\documentclass[fleqn]{lmcs}
\makeatletter
\def\@mathmargin{\parindent}
\makeatother
\pdfoutput=1
\usepackage[utf8]{inputenc}

\usepackage{lastpage}
\lmcsdoi{19}{4}{9}
\lmcsheading{}{\pageref{LastPage}}{}{}%
{Jul.~11,~2022}{Nov.~03,~2023}{}

\usepackage[british]{babel}
\usepackage{graphicx,microtype,amsfonts,amssymb,mathdefs,array,mathrsfs,amsthm,relsize}
\usepackage{hyperref}

\makeatletter
\renewcommand{\@seccntformat}[1]{%
  \ifnum\pdfstrcmp{#1}{subsection}=0
    \textbf{\csname the#1\endcsname.}
  \else
    \csname the#1\endcsname.
  \fi
}
\makeatother

\let\sc\scshape

\makeatletter
\newcommand\m@thsm@ller[2]{\mbox{\relscale{0.91}$\m@th#1#2$}}
\let\smaller\undefined
\DeclareRobustCommand\smaller[1]{\relax\ifmmode{\mathpalette\m@thsm@ller{#1}}\else{\relscale{0.91}#1}\fi}
\makeatother

\newlist{enuma}{enumerate}{10}
\setlist[enuma]{label={\normalfont(\alph*)}}
\newlist{enumr}{enumerate}{10}
\setlist[enumr]{label={\normalfont(\roman*)}}
\newlist{enum1}{enumerate}{10}
\setlist[enum1]{label={\normalfont(\arabic*)}}

\DeclareRobustCommand*{\dom}{\qopname\relax o{dom}}
\DeclareRobustCommand*{\rng}{\qopname\relax o{rng}}
\newcommand*{\id}{\mathrm{id}}
\newcommand*{\Pos}{\mathsf{Pos}}
\newcommand*{\Set}{\mathsf{Set}}
\newcommand*{\Alg}{\mathsf{Alg}}
\newcommand*{\Free}{\mathsf{Free}}
\newcommand*{\sh}{\mathrm{sh}}
\newcommand*{\Flat}{\mathrm{flat}}
\newcommand*{\sing}{\mathrm{sing}}
\newcommand*{\dist}{\mathrm{dist}}
\newcommand*{\union}{\mathrm{union}}
\newcommand*{\pt}{\mathrm{pt}}
\newcommand*{\un}{\mathrm{un}}
\newcommand*{\re}{\mathrm{re}}
\newcommand*{\comp}{\mathrm{comp}}
\newcommand*{\IN}{\mathrm{in}}
\newcommand*{\dun}{\mathrm{dun}}
\newcommand*{\sel}{\mathrm{sel}}
\newcommand*{\Id}{\mathrm{Id}}
\newcommand*{\one}{\textsf{1}}
\newcommand*{\lsem}{[\![}
\newcommand*{\rsem}{]\!]}

\newcommand*{\emptyseq}{\smaller{\langle\rangle}}
\newcommand*{\?}{\kern .08em}
\DeclareRobustCommand*{\Belowseg}{\mathord\Downarrow}
\DeclareRobustCommand*{\Aboveseg}{\mathord\Uparrow}

\newcommand\upqed{\vskip-\baselineskip\vskip-\belowdisplayskip}
\newcommand\markenddef{\hfill$\lrcorner$}

\keywords{tree algebras, power-set functor, distributive law.}

\begin{document}
\title{The Power-Set Construction for Tree Algebras}
\author[A. Blumensath]{Achim Blumensath}
\address{Masaryk University Brno}
\email{blumens@fi.muni.cz}

\begin{abstract}\noindent
We study power-set operations on classes of trees and tree algebras.
Our main result consists of a distributive law between the tree monad and the upwards-closed
power-set monad, in the case where all trees are assumed to be \emph{linear.}
For \emph{non-linear} ones, we prove that such a distributive law does not exist.
\end{abstract}

\maketitle

\section{Introduction}   

The main approaches to formal language theory are based on automata, logic, and algebra.
Each comes with their own strengths and weaknesses
and thereby complements the other two.
In the present article we focus on the algebraic approach,
which is well-known for producing proofs that are often simpler than automaton-based ones,
if not as elementary and at the cost of yielding worse complexity bounds.
Algebraic methods are especially successful at deriving structural results about
classes of languages. In particular, they are the method of choice when
deriving characterisations of subclasses of regular languages.
A~prominent example of such a result is the Theorem of Sch\"utzenberger~\cite{Schutzenberger65}
stating that a language is first-order definable if, and only if, its syntactic monoid
is aperiodic.
By now algebraic language theory is well-developed for a wide variety of settings
and types of languages, including finite words, infinite words, and finite trees.

In recent years several groups have started to work on a category-theoretic unification
of algebraic language theory
\cite{Bojanczyk15,UrbatChAdMi2017,Bojanczyk20,Blumensath20,Blumensath21}.
The motivations include both the wish to simplify the existing theories
and the need to generalise them to new settings, like infinite trees or data words.
Here, we are interested in the case of languages of infinite trees,
where an algebraic language theory has so far been missing.
We continue the technical development of the framework presented
in~\cite{Blumensath20,Blumensath21} by integrating a power-set operation.
(To be precise, we use the upwards-closed power set since our framework
is based on ordered sets.)
Such an operation has numerous uses in language theory\?:
for instance, when introducing regular expressions,
for determinisation, or when proving closure under projections.
We will present two such applications in Sections
\ref{Sect:substitutions}~and~\ref{Sect:regex} below.

There are several ways to formalise languages of infinite trees.
Most of the choices involved do not make much of a difference,
but we isolate one design choice that does\?:
a framework built on linear trees is much better behaved
than one using possibly non-linear ones.
This continues a trend already established in~\cite{Blumensath21}
indicating that non-linear trees are more complicated than linear ones.

The main technical result needed for an integration of the power-set operation is a theorem
stating that this operation can be lifted to the category of algebras under consideration.
In category-theoretical lingo this means we have to establish a distributive law
between the power-set monad and the monad our algebras are based on.
Note that there has been recent renewed interest in distributive laws also
in other parts of category theory (see, e.g., \cite{GoyPetrisanAiguier21,ZwartMarsden22}),
but the focus there is on different settings and, in particular, different functors.

We start in Section~\ref{Sect:monads} by presenting our category-theoretical
framework for infinite trees. Furthermore, we define the power-set operation we will be
investigating, and we recall the notion of a distributive law, which will be central to our work.
Section~\ref{Sect:polynomial} contains a general derivation of such laws
for a certain kind of polynomial monad, including the monad for linear trees,
and a proof that the same is not possible for non-linear trees.
The heart of the article is Section~\ref{Sect:non-linear trees}
where we will derive a partial result for non-linear trees that sometimes can be used as a
substitute for a full distributive law.
Finally, Sections \ref{Sect:substitutions}~and~\ref{Sect:regex} contain two applications\?:
the first one is a simplified proof of a recently published result on substitutions for tree
languages\?; while the second one describes how regular expressions can be defined using
power sets of non-linear trees.

\section{Monads for trees}   
\label{Sect:monads}

In algebraic language theory one uses tools from algebra to study sets~$K$ of labelled objects.
In the monadic framework from~\cite{Blumensath20,Blumensath21} these take the form
$K \subseteq \bbM\Sigma$ where $\Sigma$~is some alphabet and~$\bbM$ is a suitable monad mapping
a given set~$X$ to a set~$\bbM X$ of $X$-labelled objects of a certain kind.
Here we are mostly interested in three such monads\?:
(i)~the monad~$\bbR$ of \emph{rooted directed graphs\?;}
(ii)~the monad~$\bbT$ of \emph{linear trees\?;}
and (iii)~the monad~$\bbT^\times$ of \emph{possibly non-linear trees.}
One of our results is that the latter two behave quite differently.

Fix a countably infinite set~$X$ of \emph{variables} and let
$\Xi$~be the set of all finite subsets of~$X$.
As in~\cite{Blumensath21}, we will be working in the category $\Pos^\Xi$,
the category of $\Xi$-sorted partial orders with monotone maps as morphisms.
Thus, the objects are families $A = (A_\xi)_{\xi\in\Xi}$ where each sort~$A_\xi$
is equipped with a partial order, and the morphisms $f : A \to B$ are families
$f = (f_\xi)_{\xi\in\Xi}$ of monotone maps $f_\xi : A_\xi \to B_\xi$.
From this point on, we will use the terms `set' and `function' as a short-hand for
'ordered $\Xi$-sorted set' and `order-preserving $\Xi$-sorted function'.
For simplicity, we will frequently identify a sorted set $A = (A_\xi)_{\xi\in\Xi}$
with its disjoint union $A = \sum_{\xi\in\Xi} A_\xi$.
Using this point of view, a morphism $f : A \to B$ corresponds to a sort-preserving
and order-preserving function between the corresponding unions.

Given a set~$A$, we consider \emph{$A$-labelled, rooted, directed graphs}
which are (possibly infinite) directed graphs
with a distinguished vertex called the \emph{root} such that every vertex is reachable
by some directed path from the root.
The edges of such graphs are labelled by elements of~$X$ and the vertices by elements of~$A$
in such a way that a vertex with label $a \in A_\xi$ has exactly one outgoing edge
for each variable $x \in \xi$ and this edge is labelled by~$x$.
If there is an edge from~$v$ to~$u$ with label~$x$, we call $u$~the \emph{$x$-successor} of~$v$.
We denote the set of vertices of a graph~$g$ by $\dom(g)$.
Usually, we identify a graph~$g$ with the function $g : \dom(g) \to A$
mapping vertices of~$g$ to their labels.
We can regard $\dom(g)$ as a set in $\Pos^\Xi$
by equipping it with the trivial order and by assigning sort~$\xi$ to a vertex~$v$
if $\xi$~is the set of labels of the edges leaving~$v$.
Then $g : \dom(g) \to A$ is sort-preserving and order-preserving.
\begin{defi}
Let $A \in \Pos^\Xi$.

(a)
For a sort $\xi\in\Xi$, we denote by $\bbR_\xi A \in \Pos$
the set of all \emph{$(A + \xi)$-labelled rooted directed graphs}~$g$
(up~to isomorphism) where
\begin{itemize}
\item the elements of $\xi$ are called \emph{variables} and have sort~$\emptyset$,
\item each variable $x \in \xi$ occurs at least once in~$g$, and
\item the root of~$g$ is \emph{not} labelled by a variable.
\end{itemize}
The ordering on~$\bbR_\xi A$ is defined componentwise\?:
\begin{align*}
  g \leq h
  \quad\defiff\quad
  \dom(g) = \dom(h) \qtextq{and}
  g(v) \leq h(v)\,, \quad\text{for all } v \in \dom(g)\,.
\end{align*}
(We assume that the ordering on~$\xi$ is just the identity.)
We set
\begin{align*}
  \bbR A := (\bbR_\xi A)_{\xi\in\Xi} \in \Pos^\Xi.
\end{align*}
If $f : A \to B$ is a function, then $\bbR f : \bbR A \to \bbR B$ is the function that
applies~$f$ to each label of the given graph (leaving the labels not in~$A$ unchanged).

(b) The \emph{flattening function} $\Flat : \bbR\bbR A \to \bbR A$ maps an
$(\bbR A + \xi)$-labelled digraph~$g$ to the $(A + \xi)$-labelled digraph $\Flat(g)$
that is obtained (see Figure~\ref{Fig:flat}) from the disjoint union of all digraphs~$g(v)$,
for $v \in \dom(g)$, by
\begin{figure}
\centering
\scalebox{1.1}{\includegraphics{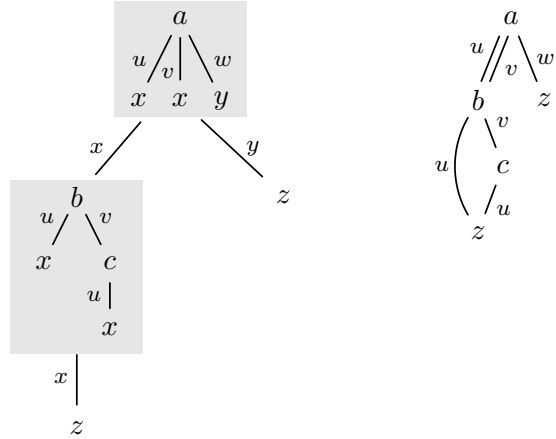}}%
\caption{The flattening operation\?: $g$ and $\Flat(g)$\label{Fig:flat}
(edge directions not shown to reduce noise)}
\end{figure}
\begin{itemize}
\item deleting from each component~$g(v)$ every vertex labelled by a variable $x \in X$ and
\item replacing every edge of $g(v)$ leading to such a vertex by an edge to the root of~$g(u_x)$,
  where $u_x$~is the $x$-successor of~$v$ in~$g$.
\end{itemize}

The \emph{singleton function} $\sing : A \to \bbR A$ maps an element $a \in A_\xi$ to
the digraph~$g$ consisting of a root labelled by~$a$ and $\abs{\xi}$~successors labelled by
the variables in~$\xi$.

(c) For $g \in \bbR A$, we denote by $\dom_0(g)$ the set of all vertices $v \in \dom(g)$
that are labelled by an element in~$A$.
\markenddef
\end{defi}

It is straightforward to check that $\bbR$~forms a monad.
(Each of the three equations can be proved by exhibiting a label-preserving bijection
between the respective domains.)
\begin{prop}
$\langle\bbR,\Flat,\sing\rangle$ forms a monad on $\Pos^\Xi$.
\end{prop}

The functors $\bbT$~and~$\bbT^\times$ can now be derived from~$\bbR$.
\begin{defi}
(a)
For a set~$A$, we denote by $\bbT^\times A \subseteq \bbR A$ the subset of all
rooted graphs that are trees, and by $\bbT A \subseteq \bbT^\times A$
the subset consisting of all trees where every variable~$x$ appears exactly once.
We call the elements of~$\bbT A$ \emph{linear trees over~$A$} and those
of $\bbT^\times A$ \emph{non-linear trees.}

For finite trees in~$\bbT^\times A$, we will frequently use the usual term notation like
\begin{align*}
  a(x,b(y,x))\,, \quad\text{for } a,b \in A\,,\ x,y \in X\,.
\end{align*}

(b)
We denote the functions $\bbT\bbT A \to \bbT A$ and $A \to \bbT A$ induced by,
respectively, $\Flat : \bbR\bbR A \to \bbR A$ and $\sing : A \to \bbR A$ also by
$\Flat$~and~$\sing$.
In cases where we want to distinguish between these versions, we add the functor as
a superscript\?: $\Flat^\bbR$, $\Flat^\bbT$, etc.

(c) We denote the category of all $\bbR$-algebras by $\Alg(\bbR)$,
and similarly for the other monads.
\markenddef
\end{defi}
The variants of $\Flat$ and $\sing$ for the functor~$\bbT^\times$ will be defined
in a later section as $\bbT^\times$~does not form a submonad of~$\bbR$.
(The family of sets~$\bbT^\times A$ is not closed under~$\Flat$.)

The fact that $\bbT$~is a monad now follows directly from the fact
that it is a restriction of~$\bbR$. To see this, we need the notion of a morphism of monads.
\begin{defi}
Let $\langle\bbM,\mu,\varepsilon\rangle$ and
$\langle\bbN,\nu,\eta\rangle$ be monads.

(a) A natural transformation $\varrho : \bbM \Rightarrow \bbN$ is a
\emph{morphism of monads} if
\begin{align*}
  \eta = \varrho \circ \varepsilon
  \qtextq{and}
  \nu \circ \varrho \circ \bbM\varrho = \varrho \circ \mu\,.
\end{align*}
In this case we say that $\bbM$~is a \emph{reduct} of~$\bbN$.

(b) Let $\varrho : \bbM \Rightarrow \bbN$ be a morphism of monads
and $\frakA = \langle A,\pi\rangle$ an $\bbN$-algebra.
The \emph{$\varrho$-reduct} of~$\frakA$ is the $\bbM$-algebra
$\frakA|_\varrho := \langle A,\pi \circ \varrho\rangle$.
If $\varrho$~is understood, we also speak of the \emph{$\bbM$-reduct} of~$\frakA$.
\markenddef
\end{defi}

The following lemma is frequently useful to prove that a functor forms a monad.
The proof is straightforward.
\begin{lem}\label{Lem: using a morphism of monads to prove monad}
Let\/ $\bbM$~and\/~$\bbN$ be functors,
$\mu : \bbM\bbM \Rightarrow \bbM$, $\nu : \bbN\bbN \Rightarrow \bbN$,
$\varepsilon : \Id \Rightarrow \bbM$, $\eta : \Id \Rightarrow \bbN$ natural transformations,
and let $\varrho : \bbM \Rightarrow \bbN$ be a natural transformation satisfying
\begin{align*}
  \eta = \varrho \circ \varepsilon
  \qtextq{and}
  \nu \circ \varrho \circ \bbM\varrho = \varrho \circ \mu\,.
\end{align*}
\begin{enuma}
\item Suppose that $\varrho$~is a monomorphism.
  If $\langle\bbN,\nu,\eta\rangle$ is a monad, then so is $\langle\bbM,\mu,\varepsilon\rangle$
  and $\varrho : \bbM \Rightarrow \bbN$ is a morphism of monads.
\item Suppose that $\varrho$~is an epimorphism and that\/ $\bbM$~preserves epimorphisms.
  If $\langle\bbM,\mu,\varepsilon\rangle$ is a monad, then so is $\langle\bbN,\nu,\eta\rangle$
  and $\varrho : \bbM \Rightarrow \bbN$ is a morphism of monads.
\end{enuma}
\end{lem}

\begin{cor}
$\langle\bbT,\Flat,\sing\rangle$ forms a monad on\/ $\Pos^\Xi$.
\end{cor}

Since our algebras are ordered it is natural to add meets (and joins) as operations.
We start by defining a monad just for meets and then add it to our algebras
via a standard construction based on so-called distributive laws.
In this and the next section we only consider the monads $\bbR$~and~$\bbT$.
The more complicated case of~$\bbT^\times$ will be dealt with separately in
Section~\ref{Sect:non-linear trees} below.

\begin{defi}
Let $A \in \Pos^\Xi$.

(a) For $X \subseteq A$, we write
\begin{align*}
  \Aboveseg X &:= \set{ a \in A }{ a \geq x \text{ for some } x \in X }\,, \\
\text{and~~} \Belowseg X &:= \set{ a \in A }{ a \leq x \text{ for some } x \in X }\,.
\end{align*}
For single elements $x \in A$, we omit the braces and simply write $\Aboveseg x$
and~$\Belowseg x$.

(b)
The \emph{(upward) power set} $\bbU A$ of~$A$ is the ordered set with domains
\begin{align*}
  \bbU_\xi A := \set{ I \subseteq A_\xi }{ I \text{ is upwards closed} }\,,
  \quad\text{for } \xi \in \Xi\,,
\end{align*}
and ordering
\begin{align*}
  I \leq J \quad\defiff\quad I \supseteq J\,,
  \quad\text{for } I,J \in \bbU_\xi A\,.
\end{align*}
For a function $f : A \to B$, we define $\bbU f : \bbU A \to \bbU B$ by
\begin{align*}
  \bbU f(I) := \Aboveseg f[I]\,,
  \quad\text{for } I \in \bbU A\,.
\end{align*}

(c)
The \emph{(downward) power set} $\bbD A$ of~$A$ is the ordered set with domains
\begin{align*}
  \bbD_\xi A := \set{ I \subseteq A_\xi }{ I \text{ is downwards closed} }\,,
  \quad\text{for } \xi \in \Xi\,,
\end{align*}
and ordering
\begin{align*}
  I \leq J \quad\defiff\quad I \subseteq J\,,
  \quad\text{for } I,J \in \bbD_\xi A\,.
\end{align*}
For a function $f : A \to B$, we define $\bbD f : \bbD A \to \bbD B$ by
\begin{align*}
  \bbD f(I) := \Belowseg f[I]\,,
  \quad\text{for } I \in \bbD A\,.
\end{align*}
\upqed
\markenddef
\end{defi}

In the following we will state and prove most results only for the functor~$\bbU$.
The case of~$\bbD$ can be handled in exactly the same way.
First, let us note that it is straightforward to check that $\bbU$~forms
a monad on~$\Pos^\Xi$.
\begin{prop}
The functor $\bbU : \Pos^\Xi \to \Pos^\Xi$ forms a monad where the multiplication
\begin{align*}
  \union : \bbU\bbU A \to \bbU A : X \mapsto \textstyle\bigcup X
\end{align*}
is given by taking the union and the singleton function
\begin{align*}
  \pt : A \to \bbU A : a \mapsto \Aboveseg\{a\}
\end{align*}
is given by the principal filter operation.
\end{prop}

\begin{exa}
The algebras for the monad~$\bbU$ are exactly those of the form $\langle A,\inf\rangle$
where $A$~is a completely ordered set.
A~function $f : A \to B$ preserves arbitrary meets if, and only if,
it is a morphism $\langle A,\inf\rangle \to \langle B,\inf\rangle$
of the corresponding $\bbU$-algebras.
The same holds for $\bbD$~and suprema.
\markenddef
\end{exa}

To show that $\bbU$~lifts to a monad on $\Alg(\bbR)$, we use a standard
technique based on distributive laws~\cite{Beck69}.
Let us recall the basic definitions and results.
\begin{defi}
Let $\langle\bbM,\mu,\varepsilon\rangle$ and $\langle\bbN,\nu,\eta\rangle$ be monads.
A natural transformation $\delta : \bbM\bbN \Rightarrow \bbN\bbM$ is a
\emph{distributive law} if
\begin{alignat*}{-1}
  \delta \circ \mu &= \bbN\mu \circ \delta \circ \bbM\delta\,,
  &\qquad \delta \circ \varepsilon &= \bbN\varepsilon\,, \\
  \delta \circ \bbM\nu &= \nu \circ \bbN\delta \circ \delta\,,
  &\qquad \delta \circ \bbM\eta &= \eta\,.
\end{alignat*}
\centering
\scalebox{1.1}{\includegraphics{Power-2.mps}}%
\vskip-\baselineskip
\markenddef
\end{defi}

We can use distributive laws to lift a monad from the base category to
the category of algebras.
\begin{defi}
Let $\langle \bbM,\mu,\varepsilon\rangle$ and $\langle \bbN,\nu,\eta\rangle$ be monads
on some category~$\calC$ and let $\bbV : \Alg(\bbM) \to \calC$ be the forgetful functor
mapping an $\bbM$-algebra to its universe.

(a)
We say that a monad $\langle\hat\bbN,\hat\nu,\hat\eta\rangle$ is a \emph{lift}
of~$\bbN$ to the category of $\bbM$-algebras if
\begin{align*}
  \bbV \circ \hat\bbN = \bbN \circ \bbV\,,\quad
  \bbV\hat\nu = \nu\,,\quad
  \bbV\hat\eta = \eta\,.
\end{align*}

(b) The \emph{Kleisli category} $\Free(\bbN)$ of~$\bbN$ is the full subcategory of~$\Alg(\bbN)$
induced by all free $\bbN$-algebras.
The \emph{free functor} $\bbF_\bbN : \calC \to \Free(\bbN)$ maps an object $C \in \calC$
to the free $\bbN$-algebra generated by~$C$, that is,
\begin{alignat*}{-1}
  \bbF_\bbN C      &:= \langle\bbN C,\nu\rangle\,,
    &&\quad\text{for objects } C \in \calC\,, \\
  \bbF_\bbN\varphi &:= \bbN\varphi\,,
    &&\quad\text{for $\calC$-morphisms } \varphi : A \to B\,.
\end{alignat*}

(c) An \emph{extension} of~$\bbM$ to $\Free(\bbN)$ is a monad
$\langle\widehat\bbM,\hat\mu,\hat\varepsilon\rangle$ on $\Free(\bbN)$
satisfying
\begin{align*}
  \widehat\bbM \circ \bbF_\bbN = \bbF_\bbN \circ \bbM\,,\quad
  \hat\mu = \bbF_\bbN\mu\,,\quad
  \hat\varepsilon = \bbF_\bbN\varepsilon\,.
\end{align*}
\upqed
\markenddef
\end{defi}
\begin{thmC}[\cite{Beck69}]\label{Thm: distributive law of monads}
Let $\langle \bbM,\mu,\varepsilon\rangle$ and $\langle \bbN,\nu,\eta\rangle$ be monads.
There exist bijections between the following objects\?:
\begin{enum1}
\item distributive laws $\delta : \bbM\bbN \Rightarrow \bbN\bbM$\?;
\item liftings\/~$\hat\bbN$ of\/~$\bbN$ to the category of\/ $\bbM$-algebras\?;
\item extensions\/~$\widehat\bbM$ of\/~$\bbM$ to the Kleisli category\/ $\Free(\bbN)$\?;
\item functions~$\kappa$ such that
  \begin{enumerate}[label={\textsc{(m\arabic*)}}]
  \item $\langle \bbN\bbM,\kappa,\eta\circ\varepsilon\rangle$ is a monad,
  \item the functions $\bbN\varepsilon$~and~$\eta$ induce morphisms of monads
    $\bbN \Rightarrow \bbN\bbM$ and $\bbM \Rightarrow \bbN\bbM$,
  \item $\kappa$~satisfies the \emph{middle unit law\?:}
    $\kappa \circ \bbN(\varepsilon \circ \eta) = \id$\,.
  \end{enumerate}
\end{enum1}
\end{thmC}

\section{Polynomial functors}   
\label{Sect:polynomial}

It is not hard to manually find a distributive law between~$\bbU$ and the monads
$\bbR$~and~$\bbT$, but it is not that much more difficult to prove a much more general result.
The monads used in language theory, including $\bbR$,~$\bbT$, and~$\bbT^\times$,
construct sets of labelled objects. The following definition captures the general form
of such a monad.
\begin{defi}
A functor $\bbF : \Pos^\Xi \to \Pos^\Xi$ is \emph{polynomial} if it is of the following
form. For objects $A \in \Pos^\Xi$,
\begin{align*}
  \bbF A = \sum_{i \in I} A^{D_i}\,,
\end{align*}
for some fixed sequence $(D_i)_{i \in I}$ of sets with $I, D_i \in \Set^\Xi$.
Hence, an element of~$\bbF A$ is of the form $\langle i,s\rangle$ with
$i \in I$ and $s : D_i \to A$ sort-preserving. The sort of $\langle i,s\rangle$ is
the sort of~$i$. We usually omit the first component from the notation and simply write~$s$.
The set $\dom(s) := D_i$ is the called \emph{domain} of~$s$.

The ordering on $\bbF A$ is defined componentwise\?:
\begin{align*}
  \langle i,s\rangle \leq \langle j,t\rangle \quad\iff\quad
  i = j \qtextq{and} s(v) \leq t(v)\,, \quad\text{for all } v \in \dom(s)\,.
\end{align*}

Finally, $\bbF$~acts on morphisms by relabelling, that is,
\begin{align*}
  \bbF f(s) := f \circ s : \dom(s) \to B\,,
  \quad\text{for } f : A \to B\,.
\end{align*}
\upqed
\markenddef
\end{defi}
\begin{rem}
(a) Note that the functors $\bbR$,~$\bbT$, and~$\bbT^\times$ are polynomial since
\begin{align*}
  \bbR A = \sum_{g \text{ graph}} A^{\dom_0(g)}\,,
\end{align*}
where the sum ranges over all countable unlabelled graphs, i.e., the set $\bbR\one$.
The same holds for the other two functors.

(b) As one can see from the above expression,
our notation for domains is not entirely consistent.
What we call $\dom(s)$ for elements of a polynomial functor,
is called $\dom_0(g)$ for graphs $g \in \bbR A$.
\markenddef
\end{rem}

As observed in~\cite{SpivakNiu21} we can describe natural transformations
between polynomial functors in the following way.
\begin{lem}\label{Lem:Yoneda for polynomial functors}
Let\/ $\bbF X = \sum_{i \in I} X^{D_i}$ and\/ $\bbG X = \sum_{j \in J} X^{E_j}$
be polynomial functors.
There exists a one-to-one correspondence between natural transformations
\begin{align*}
  \alpha : \bbF \Rightarrow \bbG
\end{align*}
and families of functions (in\/~$\Set^\Xi$)
\begin{align*}
  \alpha' : I \to J
  \qtextq{and}
  \alpha''_i : E_{\alpha'(i)} \to D_i\,,
  \quad\text{for } i \in I\,.
\end{align*}
This correspondence is given by the equation
\begin{align*}
  \alpha(\langle i,s\rangle) = \langle \alpha'(i),t\rangle
  \qtextq{with}
  t(v) = s(\alpha''_i(v))\,,
  \quad\text{for } v \in E_{\alpha'(i)}\,.
\end{align*}
\end{lem}
\begin{proof}
The above equations induce a function mapping $\alpha',\alpha''_i$ to~$\alpha$.
This function is clearly injective. Hence, it remains to show surjectivity.
Let $\alpha : \bbF \Rightarrow \bbG$ be a natural transformation.
We start by recovering the function $\alpha' : I \to J$.
Let $\one$ be a set with exactly~$1$ element~$*_\xi$ of each sort~$\xi$.
Then $\one^{D_i} = \one^{E_j}$ is a $1$-element set.
Hence, there are bijections between $\bbF\one$~and~$I$ and between $\bbG\one$~and~$J$.
In particular, the component $\alpha_\one : \bbF\one \to \bbG\one$ of~$\alpha$
induces a function $\alpha' : I \to J$.
Given some set~$A$, let $u : A \to \one$ be the unique function.
For $\langle i,s\rangle \in \bbF A$ it follows that
\begin{align*}
  \bbG u(\alpha_A(\langle i,s\rangle))
  = \alpha_\one(\bbF u(\langle i,s\rangle))
  = \alpha_\one(\langle i,{*_\xi}\rangle)
  = \langle \alpha'(i),{*_\xi}\rangle\,,
\end{align*}
where $\xi$~is the sort of~$\langle i,s\rangle$.
This implies that
\begin{align*}
  \alpha_A(\langle i,s\rangle) = \langle \alpha'(i),t\rangle\,,
  \quad\text{for some } t : E_{\alpha'(i)} \to A\,.
\end{align*}
It thus remains to construct the functions $\alpha''_i : E_{\alpha'(i)} \to D_i$.
We have just shown that $\alpha : \bbF \Rightarrow \bbG$ induces a natural transformation
$X^{D_i} \Rightarrow X^{E_{\alpha'(i)}}$.
It is therefore sufficient to show that every natural transformation
$\beta : X^D \Rightarrow X^E$ (in~$\Pos$) corresponds to a function $\beta'' : E \to D$
(in~$\Set^\Xi$) such that
\begin{align*}
  \beta(s) = t \qtextq{where} t(v) = s(\beta''(v))\,.
\end{align*}
We set
\begin{align*}
  \beta'' := \beta_D(\id_D) \in D^E.
\end{align*}
Given $s \in A^D$ and $v \in E$, it then follows that
\begin{align*}
  \beta_A(s)(v)
  = \beta_A(s^D(\id))(v)
  = s^E(\beta_D(\id))(v)
  = s^E(\beta'')(v)
  = s(\beta''(v))\,,
\end{align*}
as desired.
\end{proof}

We will need the following notation for relations between elements of
polynomial functors.
\begin{defi}
Let $\bbF : \Pos^\Xi \to \Pos^\Xi$ be a functor, $A,B$~sets,
and $p : A \times B \to A$ and $q : A \times B \to B$ the two projections.

(a) The \emph{lift} of a relation $\theta \subseteq A \times B$ is the relation
$\theta^\bbF \subseteq \bbF A \times \bbF B$ defined by
\begin{align*}
  s \mathrel\theta^\bbF t
  \quad\defiff\quad
  \bbF p(u) = s \text{ and } \bbF q(u) = t\,, \quad\text{for some } u \in \bbF\theta\,.
\end{align*}

(b) We set ${\simeq_\sh} := \theta^\bbF$ for $\theta := A \times B$.
If $s \simeq_\sh t$, we say that $s$~and~$t$ have \emph{the same shape.}

\markenddef
\end{defi}
\begin{rem}
(a)
For a polynomial functor~$\bbF X = \sum_{i \in I} X^{D_i}$
and $s \in \bbF A$, $t \in \bbF B$, we have
\begin{align*}
  s \simeq_\sh t
  \quad\iff\quad
  s \in A^{D_i} \qtextq{and} t \in B^{D_i}, \quad\text{for the same index } i \in I\,.
\end{align*}
(This implies that $s$~and~$t$ have the same sort, namely that of~$i$.)
Then
\begin{align*}
  s \mathrel\theta^\bbF t
  \quad\iff\quad
  s \simeq_\sh t
  \qtextq{and}
  s(v) \mathrel\theta t(v)\,,
  \quad\text{for all } v \in \dom(s) = \dom(t)\,.
\end{align*}

(b) In particular, two graphs $g,h \in \bbR A$ have the same shape if they have the
same underlying graph, the same sort, and the same labelling with variables.
Only the labelling with elements of~$A$ may differ.
\markenddef
\end{rem}

The goal of this section is to derive a distributive law between certain polynomial functors
and the monad~$\bbU$. Our proof closely follows similar work
from~\cite{Jacobs04,GoyPetrisanAiguier21,BojanczykKlinSalamanca22}.
The differences are mainly technical and immaterial.
The only part of the following that can be considered original seems to be
\begin{itemize}
\item the notion of linearity in Definition~\ref{Def: linear monad},
\item Theorem~\ref{Thm: dist unique}, which states that the distributive law we present is
  unique, and
\item Theorem~\ref{Thm: dist iff linear}, which states that there is no distributive law
  for non-linear monads.
\end{itemize}

Our existence proof is based on the characterisation in terms of extensions
to the Kleisli category.
We start by developing a few tools to construct such extensions.
The first observation is that we can reduce the number of conditions we have to check.
\begin{lem}\label{Lem: conditions for extension}
Let $\langle\bbM,\mu,\varepsilon\rangle$ and $\langle\bbN,\nu,\eta\rangle$ be monads on~$\calC$
and let\/ $\widehat\bbM : \Free(\bbN) \to \Free(\bbN)$ be a functor satisfying
\begin{itemize}
\item $\widehat\bbM \circ \bbF_\bbN = \bbF_\bbN \circ \bbM\,,$
\item $\bbF_\bbN\mu \circ \widehat\bbM\widehat\bbM\varphi
       = \widehat\bbM\varphi \circ \bbF_\bbN\mu\,,
       \quad\text{for every morphism } \varphi : A \to B \text{ of\/ } \Free(\bbN)\,,$
\item $\bbF_\bbN\varepsilon \circ \varphi = \widehat\bbM\varphi \circ \bbF_\bbN\varepsilon\,,
       \quad\text{for every morphism } \varphi : A \to B \text{ of\/ } \Free(\bbN)\,,$
\end{itemize}
then $\langle\widehat\bbM,\bbF_\bbN\mu,\bbF_\bbN\varepsilon\rangle$ is an extension
of $\langle\bbM,\mu,\varepsilon\rangle$ to\/ $\Free(\bbN)$.
\end{lem}
\begin{proof}
Our assumptions immediately imply that
\begin{align*}
  \bbF_\bbN\mu : \widehat\bbM\widehat\bbM \Rightarrow \widehat\bbM
  \qtextq{and}
  \bbF_\bbN\varepsilon : \mathrm{Id} \Rightarrow \widehat\bbM
\end{align*}
are natural transformations. Hence, we only have to check the monad laws for
$\langle\widehat\bbM,\bbF_\bbN\mu,\bbF_\bbN\varepsilon\rangle$.
\begin{align*}
  \bbF_\bbN\mu \circ \widehat\bbM\bbF_\bbN\mu
  &= \bbF_\bbN\mu \circ \bbF_\bbN\bbM\mu
   = \bbF_\bbN(\mu \circ \bbM\mu)
   = \bbF_\bbN(\mu \circ \mu)
   = \bbF_\bbN\mu \circ \bbF_\bbN\mu\,, \\
  \bbF_\bbN\mu \circ \bbF_\bbN\varepsilon
  &= \bbF_\bbN(\mu \circ \varepsilon)
   = \id\,, \\
  \bbF_\bbN\mu \circ \widehat\bbM\bbF_\bbN\varepsilon
  &=\bbF_\bbN\mu \circ \bbF_\bbN\bbM\varepsilon
   = \bbF_\bbN(\mu \circ \bbM\varepsilon)
   = \id\,.
\end{align*}
\upqed
\end{proof}

Note that the action of $\widehat\bbM$ on objects is already completely determined
by the requirement that $\widehat\bbM \circ \bbF_\bbN = \bbF_\bbN \circ \bbM$.
Hence, we only have to find a suitable definition of~$\widehat\bbM$ on morphisms
$\varphi : \bbN A \to \bbN B$.
For the functor $\bbN = \bbU$, we adapt a construction
from~\cite{Jacobs04,Garner20,GoyPetrisanAiguier21} based on the category of relations.
Note that every morphism $\varphi : \bbU A \to \bbU B$ of $\bbU$-algebras
is uniquely determined by its restriction $f : A \to \bbU B$ to~$A$.
The key idea is to use the following encoding of such functions.
\begin{defi}
(a) We denote the (sort-wise) power set of $A \in \Pos^\Xi$ by
$\PSet(A) \in \Pos^\Xi$.

(b) A \emph{span} is a pair of morphisms $A \leftarrow^p R \to^q B$
with the same domain.
We call a span $A \leftarrow^p R \to^q B$ \emph{injective} if
\begin{align*}
  p(c) = p(c') \qtextq{and} q(c) = q(c')
  \qtextq{implies}
  c = c'\,,
\end{align*}
and we call it \emph{closed} if, for all $a \in A$, $b \in B$, and $c \in R$,
\begin{alignat*}{-1}
  a &\leq p(c) &&\qtextq{implies} &p^{-1}(a) &\cap q^{-1}(q(c)) &&\neq \emptyset\,, \\
  b &\geq q(c) &&\qtextq{implies} &q^{-1}(b) &\cap p^{-1}(p(c)) &&\neq \emptyset\,.
\end{alignat*}

(c) The function $f : A \to \PSet(B)$ (not necessarily monotone) \emph{represented}
by a span $A \leftarrow^p R \to^q B$ is given by
\begin{align*}
  f(a) := q[p^{-1}(a)]\,,
  \quad\text{for } a \in A\,.
\end{align*}

(d)
The \emph{graph} of a function $f : A \to \bbU B$ is the relation
\begin{align*}
  G(f) := \set{ \langle a,b\rangle \in A \times B }{ b \in f(a) }\,,
\end{align*}
and the \emph{representation} of~$f$ is the span $A \leftarrow G(f) \to B$
consisting of the two projections.
\markenddef
\end{defi}

\begin{lem}
The correspondence between a function $A \to \bbU B$ and its representation forms a bijection
between \itm{\textsc{(i)}}~the set of all functions $A \to \bbU B$ in $\Pos^\Xi$ and
\itm{\textsc{(ii)}}~the set of all spans $A \leftarrow R \to B$ that are injective and closed.
\end{lem}
\begin{proof}
Let $f : A \to \bbU B$ be a function with representation $A \leftarrow^p G(f) \to^q B$.
This span is injective as every pair is uniquely determined by the values of its two components.
To see that it is also closed, suppose that $b \geq q(c)$,
for some $b \in B$ and $c \in G(f)$.
By definition of~$G(f)$, we have $c = \langle a',b'\rangle$ with $b' \in f(a')$.
As $f(a')$~is upwards closed, $b \geq q(c) = b'$ implies $b \in f(a')$.
Hence, $\langle a',b\rangle \in G(f)$ and
\begin{align*}
  \langle a',b\rangle \in q^{-1}(b) \cap p^{-1}(a')
  = q^{-1}(b) \cap p^{-1}[p(c)] \neq \emptyset\,.
\end{align*}
Similarly, suppose that $a \leq p(c)$.
Then $c = \langle a',b'\rangle$ with $b' \in f(a')$ and $a \leq a'$.
As $f$~is monotone, it follows that $f(a) \supseteq f(a')$. In particular, $b' \in f(a)$.
Hence, $\langle a,b'\rangle \in G(f)$ and
\begin{align*}
  \langle a,b'\rangle \in p^{-1}(a) \cap q^{-1}(b')
  = p^{-1}(a) \cap q^{-1}(q(c)) \neq \emptyset\,.
\end{align*}

Conversely, consider an injective, closed span $A \leftarrow^p R \to^q B$
and let $f : A \to \PSet(B)$ be the function it represents.
We have to show that $f$~is monotone and that $f(a)$~is upwards closed, for each $a \in A$.
For monotonicity, let $a \leq a'$ and $b' \in f(a')$.
We have to show that $b' \in f(a)$. By definition of~$f$, there is some $c \in R$
with $p(c) = a'$ and $q(c) = b'$.
Then $a \leq p(c)$ implies that there is some $d \in p^{-1}(a) \cap q^{-1}(q(c))$.
Consequently, $b' = q(c) = q(d) \in q[p^{-1}(a)] = f(a)$.

To show that $f(a)$~is upwards closed, suppose that $b \geq b' \in f(a) = q[p^{-1}(a)]$.
Then we can find some element $c \in R$ with $p(c) = a$ and $q(c) = b'$.
Hence, $b \geq q(c)$ and closedness implies that
we can find some element $c' \in q^{-1}(b) \cap p^{-1}[p(c)]$.
It follows that $q(c') = b$ and $p(c') = p(c) = a$.
Consequently, $b \in q[p^{-1}(a)] = f(a)$.

To conclude the proof, we have to show that these two operations
are inverse to each other.
Given a function $f : A \to \bbU B$, let $g$~be the function
represented by $A \leftarrow^p G(f) \to^q B$.
Then
\begin{align*}
  g(a) = q[p^{-1}(a)]
       = \set{ b }{ \langle a,b\rangle \in G(f) }
       = \set{ b }{ b \in f(a) }
       = f(a)\,.
\end{align*}
Conversely, consider an injective, closed span $A \leftarrow^p R \to^q B$,
let $f : A \to \bbU B$ be the function it represents,
and let $A \leftarrow^u G(f) \to^v B$ be the representation of~$f$.
Then
\begin{align*}
  G(f) &= \set{ \langle a,b\rangle }{ b \in f(a) } \\
       &= \set{ \langle a,b\rangle }{ b \in q[p^{-1}(a)] } \\
       &= \set{ \langle a,b\rangle }{ c \in R\,,\ b = q(c)\,,\ p(c) = a } \\
       &= \set{ \langle p(c),q(c)\rangle }{ c \in R }\,.
\end{align*}
Since the span $A \leftarrow^p R \to^q B$ is injective, it follows that
the function $\langle p,q\rangle : R \to G(f)$ is a bijection that commutes
with the two projections. Thus, the two spans $A \leftarrow^p R \to^q B$ and
$A \leftarrow^u G(f) \to^v B$ are isomorphic.
\end{proof}

We can compose spans by performing a pullback.
\begin{lem}\label{Lem: representation of a composition}
Let $f : A \to \bbU B$ and $g : B \to \bbU C$ be represented by,
respectively, $A \leftarrow^p R \to^q B$ and $B \leftarrow^u S \to^v C$.
Then the function
\begin{align*}
  \union \circ \bbU g \circ f : A \to \bbU C
\end{align*}
is represented by $A \leftarrow^{p\circ k} T \to^{v \circ l} C$, where
$R \leftarrow^k T \to^l S$ is the pullback of $R \to^q B \leftarrow^u S$.

\medskip
\noindent\centering
\scalebox{1.1}
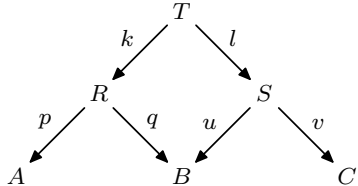
{\includegraphics{Power-3.mps}}%
%
%
%
%
%
\end{lem}
\begin{proof}
Note that the pullback in~$\Pos^\Xi$ is given by
\begin{align*}
  T = \set{ \langle r,s\rangle }{ q(r) = u(s) }
\end{align*}
and $k$~and~$l$ are the respective projections.
For $a \in A$, we therefore have
\begin{align*}
  (\union \circ \bbU g \circ f)(a)
  &= \bigcup {\set{ g(b) }{ b \in f(a) }} \\
  &= \bigcup {\bigset{ v[u^{-1}(b)] }{ b \in q[p^{-1}(a)] }} \\
  &= \bigset{ c \in C }{ c = v(s)\,,\ u(s) = q(r)\,,\ p(r) = a } \\
  &= \bigset{ c \in C }{ c = v(s)\,,\ \langle r,s\rangle \in T\,,\ p(r) = a } \\
  &= \bigset{ v(l(t)) }{ t \in T\,,\ p(k(t)) = a } \\
  &= (v \circ l)[(p \circ k)^{-1}(a)]\,.
\end{align*}
\upqed
\end{proof}

It remains to prove that polynomial functors satisfy the conditions in
Lemma~\ref{Lem: conditions for extension}.
We start by taking a look at how such a functor operates on spans.
\begin{lem}\label{Lem: polynomial functors preserve spans}
Let $\bbM : \Pos^\Xi \to \Pos^\Xi$ be a polynomial functor.
\begin{enuma}
\item $\bbM$~preserves injective and closed spans.
\item $\bbM$~preserves pullbacks.
\item $s \simeq_\sh \bbM f(s)\,, \quad\text{for all } s \in \bbM A\,,\ f : A \to B\,.$
\end{enuma}
\end{lem}
\begin{proof}
(a)
Let $A \leftarrow^p R \to^q B$ be injective and closed
and let $\bbM A \leftarrow^{\bbM p} \bbM R \to^{\bbM q} \bbM B$ be its image under~$\bbM$.

For injectivity, consider elements $s,t \in \bbM R$.
Then
\begin{align*}
  &\bbM p(s) = \bbM p(t) \qtextq{and} \bbM q(s) = \bbM q(t) \\
{}\Rightarrow\quad&
  p(s(v)) = p(t(v)) \qtextq{and} q(s(v)) = q(t(v))\,, \quad\text{for all } v\,, \\
{}\Rightarrow\quad&
  s(v) = t(v)\,, \quad\text{for all } v\,, \\
{}\Rightarrow\quad&
  s = t\,.
\end{align*}

For closedness, suppose that $s \geq \bbM q(t)$. Then
\begin{align*}
  s(v) \geq q(t(v))\,, \quad\text{for all } v\,.
\end{align*}
Hence, we can fix elements $c_v \in q^{-1}[s(v)] \cap p^{-1}[p(t(v))]$.
Setting $t'(v) := c_v$, it follows that $t' \in \bbM R$ and
\begin{align*}
  &q(t'(v)) = s(v) \qtextq{and} p(t'(v)) = p(t(v))\,, \quad\text{for all } v\,, \\
{}\Rightarrow\quad&
  \bbM q(t') = s \qtextq{and} \bbM p(t') = \bbM p(t) \\
{}\Rightarrow\quad&
  (\bbM q)^{-1}(s) \cap (\bbM p)^{-1}[\bbM p(t)] \neq \emptyset\,.
\end{align*}

Similarly, suppose that $s \leq \bbM p(t)$. Then $s(v) \leq p(t(v))$, for all~$v$.
Hence, we can fix elements $c_v \in p^{-1}[s(v)] \cap q^{-1}[q(t(v))]$.
Setting $t'(v) := c_v$, it follows that $t' \in \bbM R$ and
\begin{align*}
  &p(t'(v)) = s(v) \qtextq{and} q(t'(v)) = q(t(v))\,, \quad\text{for all } v\,, \\
{}\Rightarrow\quad&
  \bbM p(t') = s \qtextq{and} \bbM q(t') = \bbM q(t) \\
{}\Rightarrow\quad&
  (\bbM p)^{-1}(s) \cap (\bbM q)^{-1}[\bbM q(t)] \neq \emptyset\,.
\end{align*}

(b) Let $A \leftarrow^p P \to^q B$ be the pullback of $A \to^f C \leftarrow^g B$.
Then
\begin{align*}
  P = \set{ \langle a,b\rangle }{ f(a) = g(b) }
\end{align*}
and $p$~and~$q$ are the respective projections.
Similarly, the pullback of $\bbM A \to^{\bbM f} \bbM C \leftarrow^{\bbM g} \bbM B$ is
\begin{align*}
  Q &:= \bigset{ \langle s,t\rangle }{ \bbM f(s) = \bbM g(t) } \\
    &= \bigset{ \langle s,t\rangle }{ f(s(v)) = g(t(v)) \text{ for all } v } \\
    &= \bigset{ \langle s,t\rangle }{ \langle s(v), t(v)\rangle \in P \text{ for all } v }\,.
\end{align*}
Consequently, the map $\langle\bbM p,\bbM q\rangle : \bbM(A \times B) \to \bbM A \times \bbM B$
induces a bijection between $\bbM P$~and~$Q$.

(c) Setting $r := s$, $p := \id$, and $q := f$, we obtain
$\bbM p(r) = s$ and $\bbM q(r) = \bbM f(s)$.
\end{proof}

\begin{lem}\label{Lem: image of span under bbM}
Let $\bbM$~be a polynomial functor.
If $A \leftarrow^p R \to^q B$ represents $f : A \to \bbU B$,
then its image $\bbM A \leftarrow^{\bbM p} \bbM R \to^{\bbM q} \bbM B$ under~$\bbM$
represents $F : \bbM A \to \bbU\bbM B$ where
\begin{align*}
  F(s) = \set{ t \in \bbM B }{ t \in^\bbM \bbM f(s) }\,.
\end{align*}
\end{lem}
\begin{proof}
We have shown in Lemma~\ref{Lem: polynomial functors preserve spans} that
polynomial functors preserve injective closed spans.
For $s \in \bbM A$, it therefore follows that
\begin{align*}
  F(s)
  &= \bbM q\bigl[(\bbM p)^{-1}(s)\bigr] \\
  &= \set{ t \in \bbM B }{ r \in \bbM R\,,\ t = \bbM q(r)\,,\ s = \bbM p(r) } \\
  &= \set{ t \in \bbM B }
         { r \in \bbM R\,,\ t(v) = q(r(v))\,,\ s(v) = p(r(v))\,, \text{ for all } v } \\
  &= \set{ t \in \bbM B }{ t(v) \in q[p^{-1}(s(v))]\,, \text{ for all } v } \\
  &= \set{ t \in \bbM B }{ t(v) \in f(s(v))\,, \text{ for all } v } \\
  &= \set{ t \in \bbM B }{ t \in^\bbM \bbM f(s) }\,.
\end{align*}
\upqed
\end{proof}

We obtain the following proof that every polynomial functor~$\bbM$ on $\Pos^\Xi$
has an extension to $\Free(\bbU)$.
\begin{prop}\label{Prop: extension for polynomial functors}
Every polynomial functor\/~$\bbM$ on\/~$\Pos^\Xi$ induces a functor\/~$\widehat\bbM$
on\/~$\Free(\bbU)$ satisfying
\begin{align*}
  \widehat\bbM \circ \bbF_\bbN = \bbF_\bbN \circ \bbM\,.
\end{align*}
This functor maps a morphism $\varphi : \bbU A \to \bbU B$ to
\begin{align*}
  \widehat\bbM\varphi(x) := \bbM q[(\bbM p)^{-1}[x]]\,,
\end{align*}
where $A \leftarrow^p R \rightarrow^q B$ is the span representing the morphism
$\varphi \circ \pt$.
\end{prop}
\begin{proof}
As we have already explained above, for objects we are forced to set
\begin{align*}
  \widehat\bbM\langle\bbU A,\union\rangle := \langle \bbU\bbM A,\union\rangle\,.
\end{align*}
For a morphism
$\varphi : \langle\bbU A,\union\rangle \to \langle\bbU B,\union\rangle$ of free $\bbU$-algebras
we define $\widehat\bbM\varphi$ as follows.
Let $A \leftarrow^p G(\varphi) \to^q B$ be the representation of
$\varphi \circ \pt : A \to \bbU B$, and let
$\hat\varphi : \bbM A \to \bbU\bbM B$ be the function represented by
the span $\bbM A \leftarrow^{\bbM p} \bbM G(\varphi) \to^{\bbM q} \bbM B$.
Then we set
\begin{align*}
  \widehat\bbM\varphi := \union \circ \bbU\hat\varphi\,.
\end{align*}
We claim that this defines the desired functor~$\widehat\bbM$.

First, let us prove that $\widehat\bbM$ is a functor $\Free(\bbU) \to \Free(\bbU)$.
Clearly, $\widehat\bbM$ maps free $\bbU$-algebras to free $\bbU$-algebras.
Furthermore, by the above definition $\widehat\bbM\varphi$~is the free extension of
$\hat\varphi : \bbM A \to \bbU\bbM B$ to a morphism $\bbU\bbM A \to \bbU\bbM B$
of $\bbU$-algebras.
Hence, we only have to show that
\begin{align*}
  \widehat\bbM(\varphi \circ \psi) = \widehat\bbM\varphi \circ \widehat\bbM\psi\,.
\end{align*}
Let $\bbM B \leftarrow^{\bbM p} \bbM G(\varphi) \to^{\bbM q} \bbM C$ and
$\bbM A \leftarrow^{\bbM u} \bbM G(\psi) \to^{\bbM v} \bbM B$ be
the representations of $\hat\varphi$~and~$\hat\psi$.
By Lemma~\ref{Lem: representation of a composition}, the morphism
\begin{align*}
  \union \circ \bbU\hat\varphi \circ \hat\psi : \bbM A \to \bbU\bbM C
\end{align*}
is then represented by
$\bbM A \leftarrow^{\bbM u\circ k} P \to^{\bbM q\circ l} \bbM C$ where
$\bbM G(\psi) \leftarrow^k P \to^l \bbM G(\varphi)$ is the pullback of
$\bbM G(\psi) \to^{\bbM v} \bbM B \leftarrow^{\bbM p} \bbM G(\varphi)$.
Since $\bbM$~preserves pullbacks, we have $P = \bbM P'$, $k = \bbM k'$,
and $k = \bbM k'$ where $G(\psi) \leftarrow^{k'} P' \to^{l'} G(\varphi)$
is the pullback of $G(\psi) \to^v B \leftarrow^p G(\varphi)$.
Furthermore, it follows by Lemma~\ref{Lem: representation of a composition} that
$A \leftarrow^{u \circ k'} P' \to^{q\circ l'} C$ represents $\varphi \circ \psi$.
Consequently, $\widehat\bbM(\varphi \circ \psi) \circ \pt$ is also represented
by $\bbM A \leftarrow^{\bbM u\circ k} P \to^{\bbM q\circ l} \bbM C$ and we have
\begin{align*}
  \widehat\bbM(\varphi \circ \psi) \circ \pt = \union \circ \bbU\hat\varphi \circ \hat\psi =
  \widehat\bbM\varphi \circ \widehat\bbM\psi \circ \pt\,.
\end{align*}
As $\widehat\bbM(\varphi \circ \psi)$ and $\widehat\bbM\varphi \circ \widehat\bbM\psi$
are morphisms of $\bbU$-algebras, which are determined by their restriction to the
range of~$\pt$, it follows that
\begin{align*}
  \widehat\bbM(\varphi \circ \psi) = \widehat\bbM\varphi \circ \widehat\bbM\psi\,.
\end{align*}

To conclude the proof, it remains to show that
$\widehat\bbM \circ \bbF_\bbU = \bbF_\bbU \circ \bbM$.
For objects $A \in \Pos^\Xi$, this is obvious from the definition.
Hence, consider a function $f : A \to B$ and set $\varphi := \bbU f$.
Let $A \leftarrow^p G(\bbU f) \to^q B$ be the span representing~$\bbU f$.
Then $\hat\varphi : \bbM A \to \bbU\bbM B$ is represented by
$\bbM A \leftarrow^{\bbM p} \bbM G(\bbU f) \to^{\bbM q} \bbM B$.
By Lemma~\ref{Lem: image of span under bbM}, it follows that
\begin{align*}
  \widehat\bbM\bbU f(I)
  &= \bigcup \bbU\hat\varphi(I) \\
  &= \bigcup \Aboveseg\set{ \hat\varphi(s) }{ s \in I } \\
  &= \bigcup \Aboveseg\bigset{ \set{ t }{ t \in^\bbM \bbM(\bbU f \circ \pt)(s) } }{ s \in I } \\
  &= \Aboveseg\bigset{ t }{ s \in I\,,\ t \in^\bbM \bbM(\pt \circ f)(s) } \\
  &= \Aboveseg\bigset{ t }{ s \in I\,,\ t(v) \in (\pt \circ f)(s(v))  \text{ for all } v } \\
  &= \Aboveseg\bigset{ t }{ s \in I\,,\ t(v) \geq f(s(v)) \text{ for all } v } \\
  &= \Aboveseg\bigset{ t }{ s \in I\,,\ t \geq \bbM f(s) } \\
  &= \Aboveseg\bigset{ \bbM f(s) }{ s \in I } \\
  &= \bbU\bbM f(I)\,.
\end{align*}
\upqed
\end{proof}

To find the desired distributive law for polynomial monads,
it remains to prove the two remaining conditions of Lemma~\ref{Lem: conditions for extension}.
To do so, we have to make additional assumptions on our monad\?:
we require that the multiplication $\bbM\bbM \Rightarrow \bbM$ does not duplicate labels.
We will call such monads \emph{linear.}
Before we can give the formal definition, we need to take a look at the special form
the multiplication morphism for a polynomial functor takes.
\begin{rem}\label{Rem: domain map for flattening}
Let $\langle\bbM,\mu,\varepsilon\rangle$ be a monad with a polynomial functor
$\bbM X = \sum_{i \in I} X^{D_i}$.
Note that the composition $\bbM \circ \bbM$ is also a polynomial functor.
A straightforward computation yields
\begin{align*}
  \bbM\bbM X = \sum_{i \in I} \sum_{g : D_i \to I} X^{\sum_{v \in D_i} \dom(g(v))}.
\end{align*}
Thus $\bbM\bbM X = \sum_{j \in J} X^{E_j}$ where
\begin{align*}
  J := \sum_{i \in I} I^{D_i}
  \qtextq{and}
  E_{\langle i,g\rangle} := \sum_{v \in D_i} D_{g(v)}\,.
\end{align*}
Note that the identity functor~$\Id$ is polynomial, since
\begin{align*}
  \Id(A) = \sum_{\xi \in \Xi} A^{\one_\xi},
\end{align*}
where $\one_\xi$~is a set with a single element, which has sort~$\xi$.
Therefore, we can apply Lemma~\ref{Lem:Yoneda for polynomial functors} to the natural
transformations $\mu : \bbM\bbM \Rightarrow \bbM$ and $\varepsilon : \Id \Rightarrow \bbM$
and we obtain induced maps
\begin{alignat*}{-1}
  \varepsilon' &: \Xi \to I\,, &\qquad
  \varepsilon''_\xi &: D_{\varepsilon'(\xi)} \to \one_\xi\,,
   &&\quad\text{for } \xi \in \Xi\,, \\
  \mu' &: J \to I\,, &\qquad
  \mu''_j &: D_{\mu'(j)} \to E_j\,,
    &&\quad\text{for } j \in J\,.
\end{alignat*}
With our conventions regarding polynomial functors, we can write the latter as
\begin{align*}
  \mu''_s : \dom(\mu(s)) \to \sum_{v \in \dom(s)} \dom(s(v))\,,
  \quad\text{for } s \in \bbM\bbM A\,.
\end{align*}
\upqed
\markenddef
\end{rem}

\begin{defi}\label{Def: linear monad}
Let $\langle\bbM,\mu,\varepsilon\rangle$ be a monad where $\bbM$~is polynomial
and let $\mu'$, $\mu''_j$, $\varepsilon'$, and $\varepsilon''_j$ be the functions
corresponding to the natural transformations
$\mu : \bbM\bbM \Rightarrow \bbM$ and $\varepsilon : \mathrm{Id} \Rightarrow \bbM$
as above.
We call $\langle\bbM,\mu,\varepsilon\rangle$ \emph{linear} if, for all indices~$j$,
the maps~$\mu''_j$ are injective and the maps~$\varepsilon''_j$ are bijective.
\markenddef
\end{defi}
\begin{exa}
The monads $\bbR$~and~$\bbT$ are linear since each vertex of $\Flat(g)$ corresponds
to exactly one vertex of exactly one component~$g(v)$.
The monad~$\bbT^\times$ (defined below) on the other hand is not linear, since its
multiplication duplicates labels\?: substituting~$b(z)$ for~$x$ in $a(x,x)$
creates two copies of~$b$.
\markenddef
\end{exa}
\begin{rem}
Concerning terminology, the notion of a linear monad is not a~priori related
to that of a linear tree. But note that a submonad~$\bbT^0$ of~$\bbT^\times$
is linear in the above sense if, and only if, it is a submonad of~$\bbT$.
\markenddef
\end{rem}

For linear monads, we can now establish the missing identities.
We start with a technical lemma.
\begin{lem}\label{Lem: shapes for linear monads}
Let $\langle\bbM,\mu,\varepsilon\rangle$ be a linear monad on\/ $\Pos^\Xi$.
\begin{enuma}
\item $s \simeq_\sh t \!\qtextq{and}\! s(v) \simeq_\sh t(v), \text{ for all } v \in \dom(s),
  \qtextq{implies} \mu(s) \simeq_\sh \nobreak\mu(t),$
  for $s \in \bbM\bbM A$ and $t \in \bbM\bbM B$.
\item $s \simeq_\sh \mu(t) \text{ implies } s = \mu(s'), \text{ for some } s' \text{ with }
  s' \simeq_\sh t \text{ and } s'(v) \simeq_\sh t(v)\,.$
\end{enuma}
\end{lem}
\begin{proof}
Let $\mu''_j : \dom(\mu(s)) \to \sum_v \dom(s(v))$ be the injective map induced by~$\mu$.

(a) Let $p^* : \bbM A \to \one$, $q^* : \bbM B \to \one$, $p : A \to \one$, and
$q : B \to \one$.
By assumption, we have
\begin{alignat*}{-1}
  \bbM p^*(s) = \bbM q^*(t)
  \qtextq{and}
  \bbM p(s(v)) = \bbM q(t(v))\,, \quad\text{for all } v\,.
\end{alignat*}
For $w \in \dom(\mu(s))$ with $\mu''_j(w) = \langle v,u\rangle$ it follows that
\begin{alignat*}{-1}
  p\bigl(\mu(s)(w)\bigr)
  = p\bigl(s(v)(u)\bigr)
  = q\bigl(t(v)(u)\bigr)
  = q\bigl(\mu(t)(w)\bigr)\,,
\end{alignat*}
as desired.

(b) Choose $s' \in \bbM\bbM A$ such that
$s' \simeq_\sh t$,
$s'(v) \simeq_\sh t(v)$, for all~$v$, and
\begin{align*}
  s'(v)(u) :=
    \begin{cases}
      s\bigl((\mu''_j)^{-1}(v,u)\bigr) &\text{if } \langle v,u\rangle \in \rng \mu''_j\,, \\
      \text{arbitrary}                 &\text{otherwise}\,.
    \end{cases}
\end{align*}
Then we have
\begin{align*}
  s(w) = s'(v)(u)\,, \quad\text{for } \mu''_j(w) = \langle v,u\rangle\,,
\end{align*}
which, by definition of~$\mu''_j$, implies that $\mu(s') = s$.
\end{proof}

\begin{lem}
Let $\langle\bbM,\mu,\varepsilon\rangle$ be a linear monad on\/ $\Pos^\Xi$,\/
$\widehat\bbM$~its extension to\/ $\Free(\bbU)$ from
Proposition~\ref{Prop: extension for polynomial functors},
and let $\varphi : \bbU A \to \bbU B$ be a morphism of free\/ $\bbU$-algebras.
\begin{enuma}
\item $\widehat\bbM\varphi \circ \bbU\varepsilon = \bbU\varepsilon \circ \varphi\,.$
\item $\widehat\bbM\varphi \circ \bbU\mu = \bbU\mu \circ \widehat\bbM\widehat\bbM\varphi\,.$
\end{enuma}
\end{lem}
\begin{proof}
(a)
Given a morphism $\varphi : \bbU A \to \bbU B$ between free $\bbU$-algebras,
set $\varphi_0 := \varphi \circ \pt$ and let $A \leftarrow^p G(\varphi_0) \to^q B$
be the span representing it.
For $I \in \bbU A$ it then follows that
\begin{align*}
  \widehat\bbM\varphi(I) := \bbM q[(\bbM p)^{-1}[I]]\,.
\end{align*}
Since $\bbM$~is linear we furthermore have
\begin{align*}
  &\varepsilon(a) \leq \varepsilon(a') \\
  \iff\quad &a = \varepsilon(a)(v) \leq \varepsilon(a')(v) = a'\,,
    \quad\text{for all } v \in \dom(\varepsilon(a)) = \{{*}\}\,, \\
  \iff\quad &a \leq a'
\end{align*}
Hence,
\begin{align*}
  \widehat\bbM\varphi(\bbU\varepsilon(I))
  &= \widehat\bbM\bigl(\Aboveseg\set{ \varepsilon(a) }{ a \in I }\bigr) \\
  &= \bbM q\bigl[(\bbM p)^{-1}\bigl[\Aboveseg\set{ \varepsilon(a) }{ a \in I }\bigr]\bigr] \\
  &= \bbM q\bigl[\bigset{ s \in \bbM G(\varphi_0) }
                        { \bbM p(s) \geq \varepsilon(a)\,,\ a \in I }\bigr] \\
  &= \bigset{ \bbM q(\varepsilon(c)) }{ \varepsilon(c) \in \bbM G(\varphi_0)\,,\ 
                                    \bbM p(\varepsilon(c)) \geq \varepsilon(a)\,,\ a \in I } \\
  &= \bigset{ \varepsilon(b) }{ \langle a',b\rangle \in G(\varphi_0)\,,\ a' \geq a\,,\ a \in I } \\
  &= \bigset{ \varepsilon(b) }{ b \in \varphi_0(a)\,,\ a \in I } \\
  &= \bbU\varepsilon(\union(\bbU\varphi_0(I))) \\
  &= \bbU\varepsilon(\varphi(I))\,.
\end{align*}

(b)
Given a morphism $\varphi : \bbU A \to \bbU B$ between free $\bbU$-algebras,
set $\varphi_0 := \varphi \circ \pt$ and let $A \leftarrow^p G(\varphi_0) \to^q B$
be the span representing it.
It then follows that
\begin{alignat*}{-1}
  \widehat\bbM\varphi(I) &:= \bbM q[(\bbM p)^{-1}[I]]\,,
  &&\quad\text{for } I \in \bbU\bbM A\,, \\
  \widehat\bbM\widehat\bbM\varphi(I) &:= \bbM\bbM q\bigl[(\bbM\bbM p)^{-1}[I]\bigr]\,,
  &&\quad\text{for } I \in \bbU\bbM\bbM A\,.
\end{alignat*}
We start by proving that, for $r \in \bbM G(\varphi_0)$ and $s \in \bbM A$,
\begin{align*}
  \bbM p(r) \geq s
  \qtextq{implies}
  \bbM p(r') = s \text{ for some } r' \leq r\,.
\end{align*}
To see this, consider a position $v \in \dom(r)$. Then
\begin{align*}
  r(v) = \langle a_v,b_v\rangle \in G(\varphi_0)
  \qtextq{and}
  s(v) = a'_v \leq a_v\,.
\end{align*}
Hence, $b_v \in f(a_v) \geq f(a'_v)$ implies $b_v \in f(a'_v)$. Setting
\begin{align*}
  r' \simeq_\sh r
  \qtextq{and}
  r'(v) := \langle a'_v,b_v\rangle\,,
\end{align*}
we obtain $r' \in \bbM G(\varphi_0)$, $r' \leq r$, and $\bbM p(r') = s$.

To conclude the proof, note that
\begin{align*}
  \bbU\mu(\widehat\bbM\widehat\bbM\varphi(I))
  &= \bbU\mu\bigl(\bbM\bbM q\bigl[(\bbM\bbM p)^{-1}[I]\bigr]\bigr) \\
  &= \bbU\mu\bigl(\bbM\bbM q\bigl[
       \set{ r \in \bbM\bbM G(\varphi_0) }{ \bbM\bbM p(r) \in I }\bigr]\bigr) \\
  &= \bbU\mu\bigl(
       \bigset{ t \in \bbM\bbM B }{ \langle s(v)(u),t(v)(u)\rangle \in G(\varphi_0)\,,\ 
                                    s \in I }\bigr) \\
  &= \Aboveseg\bigset{ \mu(t) }
                     { r(v)(u) \in G(\varphi_0)\,,\ r(v)(u) = \langle s(v)(u), t(v)(u)\rangle\,,\
                       s \in I } \\
  &= \Aboveseg\bigset{ \mu(t) }
                     { r\in \bbM\bbM G(\varphi_0)\,,\ \bbM\bbM p(r) = s\,,\ \bbM\bbM q(r) = t\,,\
                       s \in I } \\
  &= \Aboveseg\bigset{ \mu(\bbM\bbM q(r)) }{ r \in \bbM\bbM G(\varphi_0)\,,\ \bbM p(\mu(r)) = \mu(s)\,,\ s \in I } \\
  &= \Aboveseg\bigset{ \bbM q(\mu(r)) }{ r \in \bbM\bbM G(\varphi_0)\,,\ \bbM p(\mu(r)) = \mu(s)\,,\ s \in I } \\
  &= \Aboveseg\bigset{ \bbM q(r') }{ r' \in \bbM G(\varphi_0)\,,\ \bbM p(r') = \mu(s)\,,\ s \in I } \\
  &= \Aboveseg\bigset{ \bbM q(r') }{ r' \in \bbM G(\varphi_0)\,,\ \bbM p(r') \geq \mu(s)\,,\ s \in I } \\
  &= \Aboveseg\bbM q\bigl[(\bbM p)^{-1}[\bbU\mu(I)]\bigr] \\
  &= \bbM q\bigl[(\bbM p)^{-1}[\bbU\mu(I)]\bigr] \\
  &= \widehat\bbM\varphi(\bbU\mu(I))\,,
\end{align*}
where we have used implicit universal quantification over $u$~and~$v$
and where the eight step follows by Lemma~\ref{Lem: shapes for linear monads}\,(b) and the
nineth step by the above claim.
\end{proof}

\begin{thm}\label{Thm: distributive law for polynomial functors}
Let\/ $\bbM$~be a linear monad on\/~$\Pos^\Xi$.
The functions $\dist_A : \bbM\bbU A \to \bbU\bbM A$ defined by
\begin{align*}
  \dist_A(t) := \set{ s \in \bbM A }{ s \in^\bbM t }
\end{align*}
form a distributive law\/ $\bbM\bbU \Rightarrow \bbU\bbM$.
\end{thm}
\begin{proof}
By (the proof of) Theorem~\ref{Thm: distributive law of monads},
we can obtain the desired distributive law from an extension~$\widehat\bbM$
of~$\bbM$ to $\Free(\bbU)$ by setting
\begin{align*}
  \delta := \bbV\widehat\bbM\id \circ \pt\,,
\end{align*}
where $\bbV : \Free(\bbU) \to \Pos^\Xi$ is the forgetful functor.
Note that the span representing the identity $\id : \bbU A \to \bbU A$
is $A \leftarrow^\id A \rightarrow^\pt \bbU A$.
For $t \in \bbM\bbU A$, it therefore follows that
\begin{align*}
  \delta(t)
  &= \widehat\bbM\id(\pt(t)) \\
  &= \bbM \id[(\bbM \pt)^{-1}[\Aboveseg\{t\}]] \\
  &= \set{ \bbM \id(s) }{ \bbM \pt(s) \geq t } \\
  &= \set{ s }{ \pt(s(v)) \subseteq t(v) \text{ for all } v } \\
  &= \set{ s }{ s(v) \in t(v) \text{ for all } v } \\
  &= \set{ s }{ s \in^\bbM t }\,.
\end{align*}
\upqed
\end{proof}
\begin{cor}\label{Cor: distributive law for R and T}
The functions\/ $\dist$ from above form distributive laws\/
$\bbT\bbU \Rightarrow \bbU\bbT$ and\/ $\bbR\bbU \Rightarrow \bbU\bbR$.
\end{cor}
\begin{rem}
The distributive law $\dist$ above was first stated in~\cite{Jacobs04}
for functors (not monads) on $\Set$ preserving weak pullbacks.
Our proof follows basically the same lines, except that we cannot use the algebra
of relations for $\Pos$, so we have to resort to direct calculations in several places.
See also \cite{GoyPetrisanAiguier21,BojanczykKlinSalamanca22} for similar arguments.
\markenddef
\end{rem}

We can strengthen this theorem in two ways\?:
\textsc{(i)}~the distributive law~$\dist$ is unique and
\textsc{(ii)}~there is no distributive law for non-linear monads.
We start with the former.
\begin{thm}\label{Thm: dist unique}
Let $\bbM$~be a polynomial monad on $\Pos^\Xi$ and
$\delta : \bbM\bbU \Rightarrow \bbU\bbM$ a distributive law.
Then $\delta = \dist$.
\end{thm}
\begin{proof}
$(\supseteq)$
Since $\delta$~is monotone, we have
\begin{align*}
  \delta(t)
  &\leq \inf {\set{ \delta(s) }{ s \geq t }} \\
  &\leq \inf {\bigset{ \delta(\bbM\pt(r)) }{ \bbM\pt(r) \geq t }} \\
  &= \inf {\bigset{ \pt(r) }{ \bbM\pt(r)(v) \geq t(v) \text{ for all } v }} \\
  &= \inf {\bigset{ \pt(r) }{ \pt(r(v)) \geq t(v) \text{ for all } v }} \\
  &= \bigcup {\bigset{ \pt(r) }{ \pt(r(v)) \subseteq t(v) \text{ for all } v }} \\
  &= \bigcup {\bigset{ \pt(r) }{ r(v) \in t(v) \text{ for all } v }} \\
  &= \Aboveseg {\set{ r }{ r \in^\bbM t }} \\
  &= \dist(t)\,.
\end{align*}

$(\subseteq)$
Suppose that $s \in \delta(t)$ for $t \in \bbM\bbU A$.
To prove that $s \in \dist(t)$ it is sufficient to show that $s(v) \in t(v)$, for all~$v$.
Hence, fix $v \in \dom(t)$ and
let $\theta : A \to [2]$ be the map with
\begin{align*}
  \theta(a) := \begin{cases}
                 1 &\text{if } a \in t(v)\,, \\
                 0 &\text{otherwise.}
               \end{cases}
\end{align*}
Then $\bbM\bbU\theta(t)(v) = \bbU\theta(t(v)) = \{1\}$.
Since $[2]$~is well-ordered, we can find some $r \in \bbM[2]$ such that
$\bbM\bbU\theta(t) = \bbM\pt(r)$.
It follows that
\begin{align*}
  \bbU\bbM\theta(\delta(t))
  = \delta(\bbM\bbU\theta(t))
  = \delta(\bbM\pt(t))
  = \pt(r)\,.
\end{align*}
Consequently,
\begin{align*}
  \theta(s(v)) = \bbM\theta(s)(v) \geq r(v) = 1
  \qtextq{implies}
  s(v) \in t(v)\,.
\end{align*}
\upqed
\end{proof}

As a consequence, we obtain the following strengthening of
Theorem~\ref{Thm: distributive law for polynomial functors}.
\begin{thm}\label{Thm: dist iff linear}
Let\/ $\langle\bbM,\mu,\varepsilon\rangle$~be a polynomial monad on $\Pos^\Xi$.
There exists a distributive law $\delta : \bbM\bbU \Rightarrow \bbU\bbM$ if, and only if,\/
$\bbM$~is linear.
\end{thm}
\begin{proof}
$(\Leftarrow)$ has already been proved in
Theorem~\ref{Thm: distributive law for polynomial functors}.

$(\Rightarrow)$
Suppose that $\bbM$~is not linear and let
$\mu'$,~$\mu''_j$, $\varepsilon'$, and~$\varepsilon''_j$ be the functions
corresponding to the natural transformations
$\mu : \bbM\bbM \Rightarrow \bbM$ and $\varepsilon : \mathrm{Id} \Rightarrow \bbM$
as in the definition of linearity.
By Theorem~\ref{Thm: dist unique}, it is sufficient to show that $\dist$~is not
a distributive law. For a contradiction, suppose otherwise.

By assumption, there is some index~$j$ such that $\mu''_j$~is not injective or
$\varepsilon''_j$~not bijective.
First, assume that $\mu''_j : D_{\mu'(j)} \to E_j$ is not injective, for some index~$j$.
Then there are two positions $u,v \in D_{\mu'(j)}$ with $\mu''_j(u) = \mu''_j(v)$.
Set $w := \mu''_j(u)$,
Let $A$~be a set with at least two elements $a$~and~$b$ of the same sort
as these positions (and trivial ordering),
and let $s \in \bbM\bbM\bbU\bbM A$ be such that $\dom(s) = E_j$,
\begin{align*}
  s(w) := \{\varepsilon(a),\varepsilon(b)\}
  \qtextq{and}
  s(x) = \{\varepsilon(c_x)\}\,, \quad\text{for all } x \neq w\,.
\end{align*}
By Theorem~\ref{Thm: distributive law of monads},
$\langle \bbU\bbM A,\bbU\mu \circ \dist\rangle$ is an $\bbM$-algebra
with product $\pi := \bbU\mu \circ \dist$.
Note that
\begin{align*}
     & \bigset{ \langle t(u), t(v)\rangle }{ t \in \dist(\mu(s)) } \\
{}={}& \bigset{ \langle t(u), t(v)\rangle }{ t \in^\bbM \mu(s) } \\
{}={}& \bigset{ \langle p,q\rangle }{ p \in \mu(s)(u),\, q \in \mu(s)(v) } \\
{}={}& \bigset{ \langle p,q\rangle }{ p,q \in s(w) } \\
{}={}& \bigl\{\langle\varepsilon(a),\varepsilon(a)\rangle,\,
              \langle\varepsilon(a),\varepsilon(b)\rangle,\,
              \langle\varepsilon(b),\varepsilon(a)\rangle,\,
              \langle\varepsilon(b),\varepsilon(b)\rangle \bigr\}\,.
\end{align*}
Similarly,
\begin{align*}
     & \bigset{ t(w) }{ t \in \dist(\bbM\pi(s)) } \\
{}={}& \bigset{ t(w) }{ t \in^\bbM \bbM\pi(s) } \\
{}={}& \bigset{ p }{ p \in \pi(s(w)) } \\
{}={}& \bigset{ p }{ p \in \bbU\mu(\dist(s(w))) } \\
{}={}& \bigl\{ \mu(\varepsilon(a)), \mu(\varepsilon(b)) \bigr\} \\
{}={}& \{a,b\}\,.
\end{align*}
Since every $t \in \dist(\mu(s))$ is of the form $t = \bbM\varepsilon(t_0)$,
for some $t_0 \in \bbM A$, it follows that
\begin{align*}
     & \bigset{ \langle t(u), t(v)\rangle }{ t \in \bbU\mu(\dist(\mu(s))) } \\
{}={}& \bigset{ \bigl\langle \mu(t(u)), \mu(t(v))\bigr\rangle }{ t \in \dist(\mu(s)) } \\
{}={}& \bigl\{\langle a,a\rangle,\,
              \langle a,b\rangle,\,
              \langle b,a\rangle,\,
              \langle b,b\rangle \bigr\}\,.
\end{align*}
But
\begin{align*}
     & \bigset{ \langle t(u), t(v)\rangle }{ t \in \bbU\mu(\dist(\bbM\pi(s))) } \\
{}={}& \bigset{ \langle t(w), t(w)\rangle }{ t \in \dist(\bbM\pi(s)) } \\
{}={}& \bigl\{\langle a,a\rangle,\, \langle b,b\rangle \bigr\}\,.
\end{align*}
Thus $\pi(\mu(s)) \neq \pi(\bbM\pi(s))$. A~contradiction.

It remains to consider the case where $\varepsilon''_j$~is not bijective, for some~$j$.
Then there is some sort~$\xi$ such that, for every element~$a$ of sort~$\xi$,
the domain $D := \dom(\varepsilon(a))$ is either empty or of size at least~$2$.
Let $A := \{a,b\}$ be a set with two elements of sort~$\xi$ and the trivial ordering.
If $D$~is empty, we set $s := \varepsilon(a)$ and $t := \varepsilon(b)$.
Then
\begin{align*}
  \dom(\varepsilon(s)) = \emptyset = \dom(\varepsilon(t))
  \qtextq{implies}
  \varepsilon(s) = \varepsilon(t)\,.
\end{align*}
Hence, $s = \mu(\varepsilon(s)) = \mu(\varepsilon(t)) = t$. A~contradiction.

Consequently, $D$~must have at least two elements and $\varepsilon(a) : D \to \{a\}$
is the constant function with value~$a$.
Note that $A \in \bbU A$ and
\begin{align*}
  \bbU\varepsilon(A)
  &= \bigl\{\varepsilon(a),\varepsilon(b)\bigr\}
   = \set{ s }{ s : D \to \{a,b\} \text{ a constant function} }\,, \\
  \dist(\varepsilon(A))
  &= \set{ s }{ s \in^\bbM \varepsilon(A) }
   = \set{ s }{ s : D \to \{a,b\} }\,.
\end{align*}
As $\abs{D} > 1$, there exist non-constant functions $D \to \{a,b\}$.
This implies that $\dist \circ \varepsilon \neq \bbU\varepsilon$,
a~violation of one of the axioms of a distributive law.
\end{proof}

\begin{rem}
(a) We did not make essential use of the fact that we are working with ordered sets.
All results of this section also hold in the category $\Set^\Xi$.

(b)
In the literature one can find many cases where there is no distributive law
between some variant of the power-set monad and some other monad. In particular,
there is no such law between the power-set monad and itself.
As a workaround there has been a lot of recent work
(see, e.g.,~\cite{Garner20,GoyPetrisanAiguier21})
on so-called \emph{weak distributive laws}
which satisfy the axioms for a distributive law, except possibly for
$\delta \circ \varepsilon = \bbN\varepsilon$.
A closer look at the proofs above reveals that our results also hold for
weak distributive laws if we replace linearity with the weaker condition
that only the functions~$\mu''_j$ are injective.
If we call such a monad \emph{weakly linear} it follows in particular that
there is a weak distributive law $\delta : \bbM\bbU \Rightarrow \bbU\bbM$
if, and only if, $\bbM$~is weakly linear.

(c)
In light of the above theorem, it is unsurprising that all known distributive
laws for variants of the power-set monad require some form of linearity,
although it is frequently expressed in terms of which equations the free
algebra satisfies, instead of using properties of the monad multiplication.

For instance, there is a distributive law~\cite{ManesMurly07} in $\Set$ between so-call
`commutative monads' (like the power-set monad) and finitary term monads (which are linear
in our sense).
Similarly, there is a distributive law~\cite{ManesMurly08} between certain monads
and quotients of finitary term monads by linear equations (i.e., term equations where every
variable appears exactly once on each side).

In~\cite{ZwartMarsden22} a variety of non-existence results for distributive laws
between quotients of finitary term monads is proved. In many of the cases,
one of the assumptions is that there is some term~$s$ satisfying the equation $s(x,\dots,x) = x$
(which is non-linear).

It seems that much of the existing theory could be unified if the results of this section
(which also apply to monads that are non-finitary) could be generalised from
linear polynomial monads to suitable `linear' quotients of such monads.
\markenddef
\end{rem}

\section{Non-linear trees}   
\label{Sect:non-linear trees}

It is time to properly define our third monad, that of non-linear trees,
and to prove its limited compatibility with the power-set monad.
Unfortunately, this turns out to be much more complex than the case of linear trees.
In fact, as we have seen in Theorem~\ref{Thm: dist iff linear},
there does not exist a distributive law between $\bbT^\times$~and~$\bbU$.
We will therefore forego distributive laws and directly prove the existence of
a lift of~$\bbU$ to the class of free $\bbT^\times$-algebras,
a partial result that is sufficient for many applications.
We start by defining the monad structure of~$\bbT^\times$.
\begin{defi}
(a) We denote the \emph{unravelling} (in the usual graph-theoretic sense)
of a graph $g \in \bbR_\xi A$ by $\un_0(g) \in \bbR_\xi A$.
That is, $\un_0(g)$ is the graph whose vertices consist of all finite paths of~$g$
that start at the root and there is an edge between two such paths if the second
one is the corresponding prolongation of the first one.

(b) We define $\Flat^\times : \bbT^\times\bbT^\times A \to \bbT^\times A$ and
$\sing^\times : A \to \bbT^\times A$ by
\begin{align*}
  \Flat^\times := \un_0 \circ \Flat
  \qtextq{and}
  \sing^\times := \sing\,.
\end{align*}
\upqed
\markenddef
\end{defi}

This gives us the desired monad structure for~$\bbT^\times$.
The proof is straightforward.
\begin{lem}
$\langle\bbT^\times,\Flat^\times,\sing^\times\rangle$ is a monad.
\end{lem}

In contrast to~$\bbT$, the monad~$\bbT^\times$ is \emph{not} a submonad of~$\bbR$.
Instead it is a \emph{quotient.}
\begin{lem}
$\un_0 : \bbR \Rightarrow \bbT^\times$ is a morphism of monads.
\end{lem}
\begin{proof}
We have to check that
\begin{align*}
  \sing^\times = \un_0 \circ \sing
  \qtextq{and}
  \Flat^\times \circ \un_0 \circ \bbR\un_0 = \un_0 \circ \Flat\,.
\end{align*}
The first equation immediately follows form the fact that $\un_0(\sing(a)) = \sing(a)$.
For the second one, note that the vertices of $\un_0(\Flat(g))$ correspond to
the finite paths of~$\Flat(g)$, while those of $\un_0(\Flat(\un_0(\bbR\un_0(g))))$
correspond to those of $\Flat(\un_0(\bbR\un_0(g)))$.
Furthermore, every path~$\alpha$ in a graph of the form $\Flat(h)$
corresponds to a path $(v_n)_n$ of~$h$ and a family of paths~$\beta_n$ of~$h(v_n)$
such that $\alpha$~can be identified with the concatenation $\beta_0\beta_1\dots$.
Finally, a path in $\un_0(h)$ is the same as a path in~$h$.
Consequently, each path of $\Flat(\un_0(\bbR\un_0(g)))$
corresponds to (i)~a path of~$g$ together with (ii)~a family of paths
in some components~$g(v)$ as above.
This correspondence induces a bijection between
\begin{align*}
  \dom(\un_0(\Flat(g)))
  \qtextq{and}
  \dom(\un_0(\Flat(\un_0(\bbR\un_0))))\,.
\end{align*}
As this bijection preserves the labelling it follows that
\begin{align*}
  \un_0(\Flat(g)) = \un_0(\Flat(\un_0(\bbR\un_0)))\,.
\end{align*}
\upqed
\end{proof}

The fact that there is no distributive law for~$\bbT^\times$ follows directly from
Theorem~\ref{Thm: dist iff linear} since $\bbT^\times$ is not linear.
This means that our main goal is unreachable.
But having a distributive law between $\bbT^\times$~and~$\bbU$ would be very useful.
For instance, it is needed when introducing regular expressions for infinite trees.
Therefore we will try to find a useable workaround, something weaker than an actual
distributive law that nevertheless covers the applications we have in mind.
The rest of this section is meant to get an overview over our options in this regard,
and to probe the dividing line between the possible and the impossible.
\begin{rem}
We have already mentioned above that, for cases where there is no distributive law,
there is the notion of a weak distributive law which often can be used instead.
Unfortunately, this does not work in our case since the problem above is
the monad multiplication, not the unit. ($\bbT^\times$~is not even weakly linear.)
\markenddef
\end{rem}

\subsection{Infinite sorts}   

We start with some technical remarks considering sorts.
Below we will need to deal with trees with infinitely many different variables,
that is, we have to work in the category $\Pos^{\PSet(X)}$ instead of $\Pos^\Xi$.
It is straightforward to extend the monads $\bbR$,~$\bbT$, and~$\bbT^\times$ to this
more general setting.
Hence, let us consider the following situation\?: we are given two sets $\Delta \subseteq \Gamma$
of sorts and a monad~$\bbM$ on $\Pos^\Gamma$. The following technical tools allow us to
translate between the associated categories $\Pos^\Delta$ and $\Pos^\Gamma$.
\begin{defi}
Let $\Delta \subseteq \Gamma$ be sets of sorts.

(a) The \emph{extension} of $A = (A_\xi)_{\xi \in \Delta} \in \Pos^\Delta$ to $\Pos^\Gamma$
is the set $A^\uparrow \in \Pos^\Gamma$ defined by
\begin{align*}
  A^\uparrow_\xi := \begin{cases}
                      A_\xi &\text{if } \xi \in \Delta\,, \\
                      \emptyset &\text{otherwise}\,.
                    \end{cases}
\end{align*}

(b) The \emph{restriction} of $A = (A_\xi)_{\xi \in \Gamma} \in \Pos^\Gamma$ to $\Pos^\Delta$
is the set $A|_\Delta := (A_\xi)_{\xi \in \Delta}$.
Similarly, for a function $f : A \to B$ in $\Pos^\Gamma$, we denote by
$f|_\Delta : A|_\Delta \to B|_\Delta$ the restriction to~$\Delta$.
Finally, for an $\bbM$-algebra $\frakA = \langle A,\pi\rangle$,
we set
\begin{align*}
  \frakA|_\Delta := \langle A|_\Delta,\pi|_\Delta \circ (\bbM i)|_\Delta\rangle\,,
\end{align*}
where $i : (A|_\Delta)^\uparrow \to A$ is the inclusion map.

(c) The \emph{restriction} of a functor $\bbM : \Pos^\Gamma \to \Pos^\Gamma$ to $\Pos^\Delta$
is the functor $\bbM|_\Delta : \Pos^\Delta \to \Pos^\Delta$ defined by
\begin{align*}
  \bbM|_\Delta A := (\bbM(A^\uparrow))|_\Delta\,.
\end{align*}
\upqed
\markenddef
\end{defi}
\begin{exa}
Let $\Delta := \{\emptyset,\{x\}\} \subseteq \Xi$, for some fixed $x \in X$.
The monad $\bbT|_\Delta$ is isomorphic to the functor
\begin{align*}
  \bbM\langle X_0,X_1\rangle = \langle X_1^*X_0 + X_1^\omega,\,X_1^+\rangle
\end{align*}
(up to renaming of the sorts for readability) whose algebras are (ordered) $\omega$-semigroups
$\langle S_0,S_1,\pi\rangle$.
The restriction $\bbM|_{\{1\}} X_1 = X_1^+$ is the monad for (ordered) semigroups,
while $\bbM|_{\{0\}} X_0 = X_0$ is just the identity monad.
Given an $\omega$-semigroup $\frakS = \langle S_0,S_1,\pi\rangle$, the corresponding
restrictions are the associated semigroup $\frakS|_{\{1\}} = \langle S_1,\pi_1\rangle$
and the set $\frakS|_{\{0\}} = \langle S_0,\id\rangle$.
\markenddef
\end{exa}

Let us quickly check that these definitions make sense.
\begin{lem}\label{Lem: reduct is an algebra}
Let $\langle\bbM,\mu,\varepsilon\rangle$ be a monad on~$\Pos^\Gamma$.
\begin{enuma}
\item $\bbM|_\Delta$~forms a monad with multiplication
  $(\mu \circ \bbM i)|_\Delta$ and unit map $\varepsilon|_\Delta$.
\item If\/ $\frakA$~is an\/ $\bbM$-algebra, then\/ $\frakA|_\Delta$~is an\/
  $\bbM|_\Delta$-algebra.
\end{enuma}
\end{lem}
\begin{proof}
To improve readability, let us denote the functor~$({-})|_\Delta$ by~$R$ and the
functor~$({-})^\uparrow$ by~$E$. Then $\bbM|_\Delta = R \circ \bbM \circ  E$.
We denote the inclusion $ER \Rightarrow \mathrm{Id}$ by~$i$
and the identity function $\mathrm{Id} \Rightarrow RE$ by~$e$.
One can show that $E \dashv R$ is an adjunction with unit~$e$ and counit~$i$,
but for our purposes it is sufficient to note that we have the following equalities
\begin{align*}
  i \circ Ee = \id \qtextq{and}
  Ri \circ e = \id\,,
\end{align*}
whose proofs are trivial.

(a) We have to check three axioms.
\begin{align*}
  R(\mu \circ \bbM i) \circ R\varepsilon
  &= R(\mu \circ \bbM i) \circ R\varepsilon \circ e\\
  &= R(\mu \circ \varepsilon \circ i) \circ e \\
  &= Ri \circ e\\
  &= \id\,, \displaybreak[0]\\[1ex]
  R(\mu \circ \bbM i) \circ \bbM|_\Delta R\varepsilon
  &= R(\mu \circ \bbM i) \circ \bbM|_\Delta(R\varepsilon \circ e)\\
  &= R\bigl(\mu \circ \bbM i \circ \bbM ER\varepsilon \circ \bbM Ee\bigr) \\
  &= R\bigl(\mu \circ \bbM(i \circ ER\varepsilon \circ Ee)\bigr) \\
  &= R\bigl(\mu \circ \bbM(\varepsilon \circ i \circ Ee)\bigr) \\
  &= R\bbM(\id \circ \bbM\id) \\
  &= \id\,, \displaybreak[0]\\[1ex]
  R(\mu \circ \bbM i) \circ R(\mu \circ \bbM i)
  &= R\bigl(\mu \circ \bbM i \circ \mu \circ \bbM i\bigr) \\
  &= R\bigl(\mu \circ \mu \circ \bbM\bbM i \circ \bbM i\bigr) \\
  &= R\bigl(\mu \circ \bbM\mu \circ \bbM(\bbM i \circ i)\bigr) \\
  &= R\bigl(\mu \circ \bbM(\mu \circ \bbM i \circ i\bigr) \\
  &= R\bigl(\mu \circ \bbM(i \circ ER(\mu \circ \bbM i))\bigr) \\
  &= R\bigl(\mu \circ \bbM i \circ \bbM ER(\mu \circ \bbM i)\bigr) \\
  &= R(\mu \circ \bbM i) \circ \bbM|_\Delta R(\mu \circ \bbM i)\,.
\end{align*}

(b)
Note that the product has the correct type since
\begin{align*}
  R(\pi \circ \bbM i) : R\bbM ER A \to RA
  \qtextq{and}
  \bbM|_\Delta(A|_\Delta) = R\bbM ER A\,.
\end{align*}
For the axioms of an $\bbM|_\Delta$-algebra, we have
\begin{align*}
  R(\pi \circ \bbM i) \circ R\varepsilon
  &= R(\pi \circ \bbM i \circ \varepsilon) \\
  &= R(\pi \circ \varepsilon \circ i) \\
  &= Ri \\
  &= \id\,, \\
  R(\pi \circ \bbM i) \circ \bbM|_\Delta R(\pi \circ \bbM i)
  &= R\bigl(\pi \circ \bbM(i \circ ER(\pi \circ \bbM i))\bigr) \\
  &= R\bigl(\pi \circ \bbM(\pi \circ \bbM i \circ i)\bigr) \\
  &= R\bigl(\pi \circ \mu \circ \bbM(\bbM i \circ i)\bigr) \\
  &= R\bigl(\pi \circ \bbM i \circ \mu \circ \bbM i\bigr) \\
  &= R(\pi \circ \bbM i) \circ R(\mu \circ \bbM i)\,.
\end{align*}
\upqed
\end{proof}

In the remainder of this section, we work in the category $\Pos^{\Xi_+}$
where $\Xi_+ := \PSet(\omega)$.
The functors $\bbR$,~$\bbT$, and~$\bbT^\times$ have canonical extensions to this
category, which we will denote by the same letters to keep notation readable.

\subsection{The action on the variables}   

The problem with finding a distributive law for~$\bbT^\times$ is that this monad is not linear.
Its multiplication contains an unravelling operation~$\un_0$ which is used to duplicate
arguments for variables appearing multiple times.
To continue we need a variant of this operation that also modifies the variables
of the given graph.
\begin{defi}\label{Def: un}
Let $g \in \bbR_\zeta A$ be a graph.

(a)
For a surjective function $\sigma : \zeta \to \xi$, we denote by ${}^\sigma g \in \bbR_\xi A$
the graph obtained from~$g$ by replacing each variable~$x$ by~$\sigma(x)$.

(b)
We set
\begin{align*}
  \un(g) := \langle\sigma,t\rangle\,,
\end{align*}
where $t$~is the tree obtained from the unravelling~$\un_0(g)$ by renaming the variables
so that each of them appears exactly once (note that this changes the sort)
and $\sigma$~is the function such that ${}^\sigma t = \un_0(g)$.
(To make this well-defined, we can fix a standard well-ordering on the domain,
say, the length-lexicographic one, and we number the variables in increasing order
with respect to this ordering, i.e., if $v_0 <_\mathrm{llex} v_1 <_{\mathrm{llex}} \dots$
is an enumeration of all vertices labelled by a variable, we set
$t(v_i) := x_i$, where $x_0,x_1,\dots$ is some fixed sequence of variables.)

(c) We denote by~$\bbT^\circ A$ the set of trees $t \in \bbT^\times A$ such that
$\un(t) = \langle\id,t\rangle$.
Let $\iota : \bbT^\circ \Rightarrow \bbT^\times$ be the inclusion.
(In actual calculations we will frequently omit~$\iota$ to keep the notation simple.)
\markenddef
\end{defi}
\begin{rem}
Note that the operation~$\un$ can introduce infinitely many different variables.
This is the reason why we have to work in $\Pos^{\Xi_+}$.
\markenddef
\end{rem}
\begin{exa}
$\un(a(x,y,x)) = \langle \sigma,a(x_0,x_1,x_2)\rangle$ where
the function~$\sigma$ maps $x_0,x_1,x_2$ to $x,y,x$.
Then ${}^\sigma a(x_0,x_1,x_2) = a(x,y,z)$.
\markenddef
\end{exa}

To make sense of the type of the above operations, we introduce the following monad
where every element is annotated by some function renaming the variables.
\begin{defi}
(a)
We define a functor $\bbX : \Pos^{\Xi_+} \to \Pos^{\Xi_+}$ as follows.
For $A \in \Pos^{\Xi_+}$, we set
\begin{align*}
  \bbX_\xi A :=
    \set{ \langle\sigma,a\rangle }
        { a \in A_\zeta\,,\ \sigma : \zeta \to \xi \text{ surjective} }\,.
\end{align*}
We define the order on~$\bbX_\xi A$ by
\begin{align*}
  \langle\sigma,a\rangle \leq \langle\tau,b\rangle
  \quad\defiff\quad
  \sigma = \tau \qtextq{and} a \leq b\,.
\end{align*}
For a morphism $f : A \to B$, we define $\bbX f : \bbX A \to \bbX B$ by
\begin{align*}
  \bbX f(\langle\sigma,a\rangle) := \langle\sigma,f(a)\rangle\,.
\end{align*}

(b) We define functions $\comp : \bbX\bbX A \to \bbX A$ and $\IN : A \to \bbX A$ by
\begin{align*}
  \comp(\langle\tau,\langle \sigma,a\rangle\rangle) :=
    \langle\tau \circ \sigma,a\rangle
  \qtextq{and}
  \IN(a) := \langle\id,a\rangle\,.
\end{align*}
\upqed
\markenddef
\end{defi}

\begin{lem}
$\langle\bbX,\comp,\IN\rangle$ and $\langle\bbT^\circ,\Flat,\sing\rangle$ are monads.
\end{lem}

The set~$\bbT^\times A$ carries a canonical structure of a $\bbX$-algebra.
\begin{defi}\label{Def: re}
For $\langle\sigma,t\rangle \in \bbX\bbT^\times A$, we define the \emph{reconstitution operation}
\begin{align*}
  \re(\langle\sigma,t\rangle) := {}^\sigma t \in \bbT^\times A\,.
\end{align*}
We denote its restriction to~$\bbX\bbT^\circ$ by
$\re_0 := \re \circ \bbX\iota : \bbX\bbT^\circ \Rightarrow \bbT^\times$.
\markenddef
\end{defi}

The unravelling operation on trees can now be formalised using the following
two natural transformations.
\begin{lem}\label{Lem: un and re for trees}
The inclusion morphism $\iota : \bbT^\circ \Rightarrow \bbT^\times$ is a morphism of monads.
The functions
\begin{align*}
  \un : \bbT^\times \Rightarrow \bbX\bbT^\circ,\quad
  \re_0 : \bbX\bbT^\circ \Rightarrow \bbT^\times,
  \qtextq{and}
  \re : \bbX\bbT^\times \Rightarrow \bbT^\times
\end{align*}
form natural transformations satisfying the following equations.
\begin{enuma}
\item $\re_0 \circ \un = \id$
\item $\un \circ \re = \comp \circ \bbX\un$
\item $\un \circ \iota = \IN$
\item $\re_0 \circ \comp = \re \circ \bbX\re_0$
\item $\Flat^\times \circ \re_0 = \re \circ \bbX(\Flat^\times \circ \iota)$
\item $\re \circ \IN = \id$
\item $\un \circ \re_0 = \id$
\end{enuma}
\end{lem}
\begin{proof}
The fact that $\iota$~is a morphism of monads is straightforward.
To see that $\un$~is natural, it is sufficient to note that
\begin{align*}
  \un(t) = \langle\sigma, s\rangle
  \quad\iff\quad
  \un(\bbT^\times f(t)) = \langle\sigma,\bbT^\circ f(s)\rangle\,,
\end{align*}
for every function $f : A \to B$.
For~$\re$, we have
\begin{align*}
  \bbT^\times f(\re(\langle\sigma,t\rangle))
  &= \bbT^\times f({}^\sigma t) \\
  &= {}^\sigma\bigl(\bbT^\times f(t)\bigr) \\
  &= \re(\langle\sigma,\bbT^\times f(t)\rangle)
   = \re(\bbX\bbT^\times f(\langle\sigma,t\rangle))\,.
\end{align*}
Since $\re_0 = \re \circ \bbX\iota$, this implies that $\re_0$~is natural as well.

(a) Note that $\re_0 \circ \un = \id$ holds since
\begin{align*}
  \un(t) = \langle\sigma, s\rangle
  \qtextq{implies}
  {}^\sigma s = t\,,
  \quad\text{for trees } t \in \bbT^\times A\,.
\end{align*}

(b) Suppose that $\un(t) = \langle\sigma, s\rangle$ and $\un({}^\tau t) = \langle\rho,r\rangle$.
Then
\begin{align*}
  {}^{\tau\circ\sigma} s = {}^\tau t = {}^\rho r\,.
\end{align*}
In particular, $s$~and~$r$ only differ in the labelling of the variables.
But $s,r \in \bbT^\circ A$ implies that the variables appear in the same order
in both trees. Hence, $s = r$ and it follows that $\tau \circ \sigma = \rho$.
Consequently,
\begin{align*}
  \un(\re(\langle\tau,t\rangle))
  &= \langle\rho,r\rangle \\
  &= \langle\tau\circ\sigma,s\rangle
   = \comp(\langle\tau,\langle\sigma,s\rangle\rangle)
   = \comp(\bbX\un(\langle\tau,t\rangle))\,.
\end{align*}

(c)--(f) We have
\begin{align*}
  \un(\iota(t)) &= \langle\id,t\rangle = \IN(t)\,, \\[1ex]
  \re_0\bigl(\comp\bigl(\langle\sigma,\langle\tau,t\rangle\rangle\bigr)\bigr)
  &= \re_0(\langle\sigma\circ\tau,t\rangle) \\
  &= {}^{\sigma\circ\tau}\iota(t) \\
  &= {}^\sigma({}^\tau\iota(t)) \\
  &= {}^\sigma \re_0(\langle\tau,t\rangle) \\
  &= \re\bigl(\bigl\langle\sigma, \re_0(\langle\tau,t\rangle)\bigr\rangle\bigr)
   = \re\bigl(\bbX\re_0\bigl(\langle\sigma, \langle\tau,t\rangle\rangle\bigr)\bigr)\,,
      \displaybreak[0]\\[1ex]
  \Flat^\times(\re_0(\langle\sigma,t\rangle))
  &= \Flat^\times({}^\sigma\iota(t)) \\
  &= {}^\sigma(\Flat^\times \circ \iota)(t) \\
  &= \re\bigl(\bigl\langle\sigma,(\Flat^\times \circ \iota)(t)\bigr\rangle\bigr)
   = (\re \circ \bbX(\Flat^\times \circ \iota))(\langle\sigma,t\rangle)\,,
      \displaybreak[0]\\[1ex]
  \re(\IN(t)) &= \re(\langle\id,t\rangle) = {}^\id t = t\,.
\end{align*}

(g) By~(c), we have
\begin{align*}
  \un \circ \re_0
    &= \un \circ \re \circ \bbX\iota
     = \comp \circ \bbX\un \circ\bbX\iota
     = \comp \circ \bbX\IN
     = \id\,.
\end{align*}
\upqed
\end{proof}
We can understand point~(a) of this lemma as saying that~$\bbT^\times$
is a retract of~$\bbX\bbT^\circ$, but only as functors, not necessarily as monads.
For the latter we first have to establish that $\bbX\bbT^\circ$ forms a monad
and that the operations $\un$~and~$\re_0$ are morphisms of monads.
\pagebreak
\begin{prop}\label{Prop: XT monad}\leavevmode
\begin{enuma}
\item $\bbX\bbT^\circ$~forms a monad with multiplication
  \begin{align*}
    \un \circ \re \circ \bbX(\Flat^\times \circ \iota \circ \bbT^\circ\re_0)
      : \bbX\bbT^\circ\bbX\bbT^\circ \Rightarrow \bbX\bbT^\circ
  \end{align*}
  and unit
  \begin{align*}
    \IN \circ \sing : \Id \Rightarrow \bbX\bbT^\circ\,.
  \end{align*}
\item $\re_0 : \bbX\bbT^\circ \Rightarrow \bbT^\times$
  and\/ $\un : \bbT^\times \Rightarrow \bbX\bbT^\circ$ are isomorphisms of monads.
\item $\IN : \bbT^\circ \Rightarrow \bbX\bbT^\circ$ is an injective morphism of monads.
\end{enuma}
\end{prop}
\begin{proof}
(a),~(b)
By Lemma~\ref{Lem: un and re for trees} (c),~(e), and~(a), we have
\begin{align*}
  \re_0 \circ \IN \circ \sing
  &= \re_0 \circ \un \circ \iota \circ \sing
   = \iota \circ \sing
   = \sing^\times, \\
  \Flat^\times \circ \re_0 \circ \bbX\bbT^\circ\re_0
  &= \re \circ \bbX(\Flat^\times \circ \iota \circ \bbT^\circ\re_0) \\
  &= \re_0 \circ \un \circ \re \circ \bbX(\Flat^\times \circ \iota \circ \bbT^\circ\re_0)\,.
\end{align*}
As~$\re_0$~is a surjective natural transformation, most of the claim therefore follows
by Lemma~\ref{Lem: using a morphism of monads to prove monad}.
It only remains to check that $\un$~is also a morphism of monads.
For this, note that by Lemma~\ref{Lem: un and re for trees} (c),~(a), and~(e) we have
\begin{align*}
  \IN \circ \sing &= \un \circ \iota \circ \sing = \un \circ \sing^\times\,, \\
  \un \circ \Flat^\times
    &= \un \circ \Flat^\times \circ \re_0 \circ \un \\
    &= \un \circ \re \circ \bbX(\Flat^\times \circ \iota) \circ \un \\
    &= \un \circ \re \circ \bbX(\Flat^\times \circ \iota) \circ \un \circ \bbT^\times(\re_0 \circ \un) \\
    &= \un \circ \re \circ \bbX(\Flat^\times \circ \iota \circ \bbT^\circ\re_0) \circ \un \circ \bbT^\times\un\,.
\end{align*}

(c) As $\un$~and~$\iota$ are morphisms of monads, so is $\un \circ \iota = \IN$.
\end{proof}

\begin{cor}
$\bbT^\times \cong \bbX\bbT^\circ$ (as monads)
\end{cor}

One could hope to construct a distributive law $\bbT^\circ\bbX \Rightarrow \bbX\bbT^\circ$
by applying the Theorem of Beck (Theorem~\ref{Thm: distributive law of monads}) to the monad
structure on~$\bbX\bbT^\circ$. This does not work for the following reason.
\begin{lem}
The natural transformation $\bbX\sing : \bbX \Rightarrow \bbX\bbT^\circ$ is \emph{not}
a morphism of monads.
\end{lem}
\begin{proof}
The following of the two axioms fails\?:
\begin{align*}
  \bbX(\Flat^\times \circ \iota \circ \bbT^\circ\re_0) \circ \bbX\sing \circ \bbX\bbX\sing
  \neq \bbX\sing \circ \comp.
\end{align*}
To see this, fix $\langle\sigma,\langle\tau,a\rangle\rangle \in \bbX\bbX A$.
Then
\begin{align*}
     & (\bbX(\Flat^\times \circ \iota \circ \bbT^\circ\re_0) \circ \bbX\sing \circ \bbX\bbX\sing)
         (\langle\sigma,\langle\tau,a\rangle\rangle) \\
{}={}& \bbX(\Flat^\times \circ \iota \circ \bbT^\circ\re_0 \circ \sing \circ \bbX\sing)
         (\langle\sigma,\langle\tau,a\rangle\rangle) \\
{}={}& \bbX(\Flat^\times \circ \iota \circ \sing \circ \re_0 \circ \bbX\sing)
         (\langle\sigma,\langle\tau,a\rangle\rangle) \\
{}={}& \bbX(\re_0 \circ \bbX\sing)(\langle\sigma,\langle\tau,a\rangle\rangle) \\
{}={}& \langle\sigma,{}^\tau\sing(a)\rangle\,,
\end{align*}
whereas
\begin{align*}
     & (\bbX\sing \circ \comp) (\langle\sigma,\langle\tau,a\rangle\rangle) \\
{}={}& \bbX\sing(\langle\sigma\circ\tau,a\rangle) \\
{}={}& \langle\sigma\circ\tau,\sing(a)\rangle\,.
\end{align*}
For $\tau \neq \id$, these two values are different.
\end{proof}

\subsection{Graphs and unravellings}   

The next step is to transfer the unravelling operation from~$\bbT^\times A$
to arbitrary sets.
\begin{defi}
(a)
An \emph{unravelling structure} $\langle A,\re,\un\rangle$
consists of a set $A \in \Pos^{\Xi_+}$ equipped with two functions
\begin{align*}
  \re : \bbX A \to A \qtextq{and} \un : A \to \bbX A
\end{align*}
such that $\langle A,\re\rangle$ forms a $\bbX$-algebra while $\un$~satisfies
\begin{align*}
  \bbX\un \circ \un = \bbX\IN \circ \un
  \qtextq{and}
  \re \circ \un = \id\,.
\end{align*}
We call $\un(a)$~the \emph{unravelling} of~$a$.
To keep notation simple, we write
\begin{align*}
  {}^\sigma a := \re(\langle\sigma, a\rangle)\,.
\end{align*}

(b) A \emph{morphism of unravelling structures} is a function $\varphi : A \to B$
satisfying
\begin{align*}
  \un \circ \varphi = \bbX\varphi \circ \un
  \qtextq{and}
  \varphi \circ \re = \re \circ \bbX\varphi\,.
\end{align*}
\upqed
\markenddef
\end{defi}

Clearly, the operations $\re$ and $\un$ defined above for trees $t \in \bbT^\times A$
induce an unravelling structure on~$\bbT^\times A$.
But note that this is not the case for~$\bbR A$ since we have
$\re(\un(g)) \neq g$, for every $g \in \bbR A$ that is not a tree.

\begin{exa}
For each $\bbT^\times$-algebra $\frakA = \langle A,\pi\rangle$,
we can equip the universe~$A$ with the \emph{trivial} unravelling structure where
\begin{align*}
  \un := \IN
  \qtextq{and}
  {}^\sigma a := \pi({}^\sigma \sing(a))\,.
\end{align*}
\upqed
\markenddef
\end{exa}

\begin{rem}
Note that the monad multiplication $\Flat^\times$ is
\emph{not} a morphism of unravelling structures since
$\un \circ \Flat^\times \neq \bbX\Flat^\times \circ \un$.
In what follows we will therefore \emph{not} work in the category of unravelling structures
and their morphisms. Instead we will work in the weaker category of unravelling structures
with arbitrary monotone maps as morphisms.
\markenddef
\end{rem}

As a technical tool, we use the following generalisation of the unravelling relation for graphs
where we do not only unravel the graph itself but also each label.
The intuition is as follows. Suppose we are given a relation $\theta \subseteq A \times B$
and a graph $h \in \bbR B$.
We construct an (unravelled) graph $g \in \bbR A$ as follows.
Starting at the root~$v$, we pick some element $c \mathrel\theta h(v)$,
and label~$g(v)$ by the unravelling of~$c$.
Then we recursively choose labellings for the successors.
Note that the shapes of $g$~and~$h$ are different since we are unravelling~$g$,
so the labels in~$h$ might have a higher arity than the corresponding ones in~$g$.
Consequently, we simultaneously construct a graph homomorphism $\varphi : g \to h$
to keep track of which vertices of~$g$ correspond to which ones of~$h$.

To simplify the definition, we will split the construction into two stages.
In the first step we apply the unravelling operation to every label of~$h$,
resulting in a graph $\bbR\un(h) \in \bbR\bbX B$.
What is then left for the second step is the following relation,
which does the choosing of the label and the unravelling of the tree.
What makes this operation complicated is the fact that the unravelling depends on the chosen
label, while the label may depend on which copy (produced by previous unravelling steps)
of a vertex we are at. So we cannot separate the second stage into two independent phases.
\begin{defi}
(a)
Let $g \in \bbR_\xi A$ and $h \in \bbR_\zeta B$.
A~\emph{graph homomorphism} is a function $\varphi : \dom(g) \to \dom(h)$ such that
\begin{itemize}
\item $\varphi$ maps the root of~$g$ to the root of~$h$\?;
\item $\varphi(u)$~is a successor of~$\varphi(v)$ if, and only if, $u$~is a successor of~$v$
  (not necessarily with the same edge labelling)\?; and
\item $\varphi(v)$~is labelled by a variable if, and only if, $v$~is labelled by one.
\end{itemize}

(b) Suppose that $\varphi : g \to h$ is a surjective graph homomorphism and
let $v \in \dom(g)$ be a vertex of sort~$\xi$ with successors $(u_x)_{x \in \xi}$
and suppose that $\varphi(v)$ has sort~$\zeta$.
We denote by $\varphi_{/v} : \xi \to \zeta$ the function such that
\begin{align*}
  \varphi(u_x) \text{ is the $\varphi_{/v}(x)$-successor of } \varphi(v)\,.
\end{align*}

(c)
Let $s \in \bbR A$, $t \in \bbR B$, and $\theta \subseteq \bbX A \times B$.
We write
\begin{align*}
  \varphi,\sigma : s \mathrel\theta^\sel t
\end{align*}
if the following conditions are satisfied.
\begin{itemize}
\item $s \in \bbT^\circ A$
\item $\varphi : s \to t$ is a surjective graph homomorphism.
\item $\sigma : \xi \to \zeta$ is surjective.
\item $\langle\varphi_{/v},s(v)\rangle \mathrel\theta t(\varphi(v))\,,%
      \quad\text{for every } v \in \dom_0(g)\,.$
\item $\rlap{\sigma(s(v)) = t(\varphi(v))\,,}%
       \hphantom{\langle\varphi_{/v},s(v)\rangle \mathrel\theta t(\varphi(v))\,,{}}%
       \quad\text{if $s(v) = x$ is a variable.}$\markenddef
\end{itemize}
\end{defi}

We are mostly interested in the cases where $\theta$~is either the identity~$=$
or set membership~$\in$. The resulting relations are
\begin{alignat*}{-1}
  \varphi,\sigma &: s =^\sel t\,,
  &&\quad\text{for } s \in \bbT^\times A \text{ and } t \in \bbT^\times\bbX A\,, \\
  \varphi,\sigma &: s \in^\sel t\,,
  &&\quad\text{for } s \in \bbT^\times A \text{ and } t \in \bbT^\times\bbU\bbX A\,.
\end{alignat*}
Combining them with the unravelling operation as explained above, we obtain the relations
\begin{alignat*}{-1}
  \varphi,\sigma &: s =^\un t
  &&\quad\defiff\quad
  &\varphi,\sigma &: s =^\sel \bbR\un(t)\,, \\
  \varphi,\sigma &: s \in^\un t\,,
  &&\quad\defiff\quad
  &\varphi,\sigma &: s \in^\sel \bbR\bbU\un(t)\,.
\end{alignat*}

\begin{exa}
We have $\varphi,\sigma : g \in^\un h$ where $g$~is the tree on the left,
$h$~the one on the right, $\varphi : g \to h$ is the obvious homomorphism, and
$\sigma : \{x,y,z\} \to \{x\}$.
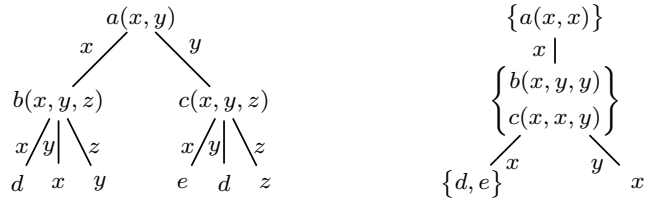
\begin{center}
\scalebox{1.1}{\includegraphics{Power-4.mps}}%
\end{center}
\upqed
\markenddef
\end{exa}
\begin{rem}
(a)
For every graph~$g$, there exists a canonical graph homomorphism $\varphi : \un_0(g) \to g$.

(b)
Note that
\begin{align*}
  \varphi,\sigma : g =^\sel k
  \qtextq{and}
  k \mathrel\theta^\bbR h
  \qtextq{implies}
  \varphi,\sigma : g \mathrel\theta^\sel h\,,
\end{align*}
but the converse is generally not true since the function~$\varphi$ does not need
to be injective and we can choose different values
$\langle\varphi_{/u},c_u\rangle,\,\langle\varphi_{/v},c_v\rangle \mathrel\theta h(w)$
for $u,v \in \varphi^{-1}(w)$.
For this reason, we cannot reduce the relation~$\in^\sel$ to the much simpler~$=^\sel$.
\markenddef
\end{rem}

Let us derive an algebraic description of the relation $\varphi,\sigma : s =^\sel t$
that is much easier to work with.
We introduce a function~$\un^+$ satisfying
\begin{align*}
  \langle\sigma,s\rangle = \un^+(t)
  \quad\iff\quad
  \varphi,\sigma : s =^\sel t\,, \quad\text{for some } \varphi\,,
\end{align*}
and a similar function~$\dun$ associated with the relation~$=^\un$.
\begin{defi}
(a) For a set~$A$, we define the \emph{strong unravelling operation}
$\un^+ : \bbT^\times\bbX A \to \bbX\bbT^\circ A$ by
\begin{align*}
  \un^+ := \un \circ \Flat^\times \circ \bbT^\times(\re_0 \circ \bbX\sing)\,.
\end{align*}

(b) For an unravelling structure~$A$, we define the \emph{deep unravelling operation}
$\dun : \bbT^\times A \to \bbX\bbT^\circ A$ by
\begin{align*}
  \dun := \un^+ \circ \bbT^\times\un\,.
\end{align*}
\upqed
\markenddef
\end{defi}
\begin{exa}
To understand the definition of~$\un^+$, let us consider the following tree
$t \in \bbT^\times\bbX A$. Below we have depicted $t$~itself, the intermediate terms
$t' := \bbT^\times(\re_0 \circ \bbX\sing)(t)$ and $t'' := \Flat(t')$, and
the end result~$\un^+(t)$.
\begin{center}
\scalebox{1.1}{\includegraphics{Power-5.mps}}%
\end{center}
Here $a,b \in A_{\{x_0,x_1\}}$, $c \in A_{\{x_0\}}$, and $\sigma_{ij}$~denotes the function
mapping $x_0 \mapsto x_i$ and $x_1 \mapsto x_j$.

\markenddef
\end{exa}

Let us check that the above definitions have the desired effect.
\begin{lem}\label{Lem: characterisation of =^un}
We have
\begin{alignat*}{-1}
  \langle\sigma,s\rangle &= \un^+(t)
  &&\quad\iff\quad
  \varphi,\sigma &: s =^\sel t\,, \quad\text{for some } \varphi\,, \\
  \langle\sigma,s\rangle &= \dun(t)
  &&\quad\iff\quad
  \varphi,\sigma &: s =^\un t\,, \quad\text{for some } \varphi\,.
\end{alignat*}
\end{lem}
\begin{proof}
We only have to prove the first equivalence. Then the second one follows by definition
of $\dun$~and~$=^\un$.
Hence, set
\begin{align*}
  r := \bbR(\re_0 \circ \bbX\sing)(t)
  \qtextq{and}
  \langle\sigma,s\rangle := \un(\Flat^\times(r))\,,
\end{align*}
let $\varphi : \dom(\Flat^\times(r)) \to \dom(t)$ be the homomorphism from above,
let $\varphi : \dom(\Flat^\times(r)) \to \dom(t)$ be the graph homomorphism induced by the
canonical map
\begin{align*}
  \dom_0(\Flat^\times(r)) \to \sum_{v \in \dom_0(r)} \dom_0(r(v))\,,
\end{align*}
and suppose that $\varphi',\sigma' : s' =^\sel t$. We have to show that
\begin{align*}
  \varphi = \varphi'\,,\quad \sigma = \sigma'\,, \qtextq{and} s = s'\,.
\end{align*}

We start by proving that $\varphi(v) = \varphi'(v)$ and $s(v) = s'(v)$, by induction on~$v$.
For the root $v = \emptyseq$ of~$\Flat^\times(r)$,
we have $\varphi(\emptyseq) = \emptyseq = \varphi'(\emptyseq)$.

For the inductive step, suppose that we have already shown that $\varphi(v) = \varphi'(v)$.
We will prove that $s(v) = s'(v)$ and that $\varphi(u) = \varphi'(u)$,
for every successor~$u$ of~$v$.
By definition of~$=^\sel$, we have
\begin{align*}
  t(\varphi'(v)) = \langle\varphi'_{/v},s'(v)\rangle\,,
  \quad\text{for } v \in \dom(s')\,.
\end{align*}
This implies that
\begin{align*}
  r(\varphi'(v))
  = (\re_0 \circ \bbX\sing)\bigl(\langle\varphi'_{/v},s'(v)\rangle\bigr)
  = {}^{\varphi'_{/v}}\sing(s'(v))\,.
\end{align*}
Consequently,
\begin{align*}
  s(v)
  = \Flat^\times(r)(v)
  = r(\varphi(v))(\emptyseq)
  = r(\varphi'(v))(\emptyseq)
  = s'(v)\,.
\end{align*}
To complete the induction, it remains to show that $\varphi_{/v} = \varphi'_{/v}$.
Let $(u_x)_x$ be the successors of~$v$ in~$s$ and
let $(w_y)_y$ be the successors of~$\varphi(v)$ in~$r$.
Then
\begin{align*}
  r(\varphi(v)) = {}^{\varphi'_{/v}}\sing(s(v))
\end{align*}
implies that the $x$-successor of~$v$ in~$s$ corresponds (via~$\varphi$)
to the $\varphi'_{/v}(x)$-successor of~$\varphi(v)$ in~$r$.
Thus
\begin{align*}
  \varphi(u_x) = w_{\varphi'_{/v}(x)}\,.
\end{align*}
But, by definition of~$\varphi_{/v}$, we also have $\varphi(u_x) = w_{\varphi_{/v}(x)}.$
Hence,
\begin{align*}
  \varphi_{/v}(x) = \varphi'_{/v}(x)\,.
\end{align*}

This completes the induction.
To finish the proof it remains to show that $\sigma = \sigma'$ and
that $s(v) = s'(v)$, for all $v \in \dom(s) \setminus \dom_0(s)$.
For the latter, note that the vertices of~$s$ carrying a variable are the same
as those of~$s'$ carrying one. Since the variable labelling is determined
by the ordering of these vertices with respect to the length-lexicographic order,
it follows that the two labellings coincide.

Hence, let~$v$ be such a vertex. Then
\begin{align*}
  \sigma(s(v)) = \Flat^\times(r)(v) = r(\varphi(v)) = t(\varphi(v)) = \sigma'(s'(v))
  = \sigma'(s(v))\,.
\end{align*}
Thus, $\sigma(x) = \sigma'(x)$, for all~$x$, which implies that $\sigma = \sigma'$
\end{proof}
Let us collect a few basic properties of the operations we have just introduced.
\begin{lem}\label{Lem: dun commutes with Flat}\leavevmode
\begin{enuma}
\item $\bbX(\un \circ \Flat^\times \circ \iota) \circ \dun =
         \bbX(\IN \circ \Flat^\times \circ \iota) \circ \dun$
\item $\Flat^\times \circ \re \circ \dun = \Flat^\times$
\item $\un \circ \Flat^\times = \bbX(\Flat^\times \circ \iota) \circ \dun$
\item $\un^+ \circ \bbT^\times\IN = \un$
\item $\un^+ \circ \sing^\times = \bbC\sing$
\end{enuma}
\end{lem}
\begin{proof}
(a)
Let $\langle\sigma,s\rangle = \dun(t)$.
By Lemma~\ref{Lem: characterisation of =^un}, it follows that $\varphi,\sigma : s =^\un t$.
Consequently, we have
\begin{align*}
  \un(t(\varphi(v))) = \langle\varphi_{/v}, s(v)\rangle\,,
  \quad\text{for all } v \in \dom_0(s)\,.
\end{align*}
In particular, $s(v) \in \bbT^\circ A$ and, therefore, $s \in \bbT^\circ\bbT^\circ A$.
This implies that $\Flat(s) \in \bbT^\circ A$.
Hence, $\un(\Flat(s)) = \langle\id,\Flat(s)\rangle$ and we have
\begin{align*}
  \bbX(\un \circ \Flat)(\dun(t))
  &= \langle\sigma,\un(\Flat(s))\rangle \\
  &= \langle\sigma,\langle\id,\Flat(s)\rangle\rangle \\
  &= \langle\sigma,\IN(\Flat(s))\rangle
   = \bbX(\IN \circ \Flat)(\dun(t))\,.
\end{align*}

(b)
From Lemma~\ref{Lem: un and re for trees} it follows that
\begin{align*}
  \Flat^\times \circ \re_0 \circ \bbX\sing \circ \un
  &= \re_0 \circ \bbX(\Flat^\times \circ \iota) \circ \bbX\sing \circ \un \\
  &= \re_0 \circ \bbX(\Flat^\times \circ \sing^\times) \circ \un \\
  &= \re_0 \circ \un \\
  &= \id\,.
\end{align*}
Consequently,
\begin{align*}
  \Flat^\times \circ \re \circ \dun
  &= \Flat^\times \circ \re \circ \un \circ \Flat^\times \circ
        \bbT^\times(\re_0 \circ \bbX\sing \circ \un) \\
  &= \Flat^\times \circ \Flat^\times \circ \bbT^\times(\re_0 \circ \bbX\sing \circ \un) \\
  &= \Flat^\times \circ \bbT^\times\Flat^\times \circ \bbT^\times(\re_0 \circ \bbX\sing \circ \un) \\
  &= \Flat^\times \circ \bbT^\times\id \\
  &= \Flat^\times.
\end{align*}

(c)
By (a)~and Lemma~\ref{Lem: un and re for trees}, we have
\begin{align*}
  \bbX(\Flat^\times \circ \iota) \circ \dun
  &= \comp \circ \bbX(\IN \circ \Flat^\times \circ \iota) \circ \dun \\
  &= \comp \circ \bbX(\IN \circ \Flat^\times \circ \iota) \circ \dun \\
  &= \comp \circ \bbX\un \circ \bbX(\Flat^\times \circ \iota) \circ \dun \\
  &= \un \circ \re \circ \bbX(\Flat^\times \circ \iota) \circ \dun \\
  &= \un \circ \Flat^\times \circ \re_0 \circ \dun \\
  &= \un \circ \Flat^\times.
\end{align*}

(d) We have
\begin{align*}
   \un^+ \circ \bbT^\times\IN
   &= \un \circ \Flat^\times \circ \bbT^\times(\re_0 \circ \bbX\sing)
        \circ \bbT^\times\IN \\
   &= \un \circ \Flat^\times \circ \bbT^\times(\re_0 \circ \IN \circ \sing) \\
   &= \un \circ \Flat^\times \circ \bbT^\times\sing \\
   &= \un\,.
\end{align*}

(e) \vspace*{-18pt}
\begin{align*}
  \un^+ \circ \sing^\times
  &= \un \circ \Flat^\times \circ \bbT^\times(\re_0 \circ \bbX\sing) \circ \sing^\times \\
  &= \un \circ \Flat^\times \circ \sing^\times \circ \re_0 \circ \bbC\sing \\
  &= \un \circ \re_0 \circ \bbX\sing \\
  &= \bbX\sing\,.
\end{align*}
\upqed
\end{proof}

In Lemma~\ref{Lem: characterisation of =^un}, we have found an algebraic
characterisation of the relations $=^\sel$~and~$=^\un$ in terms of the
operations $\un^+$~and~$\dun$.
Unfortunately, there does not seem to be a purely algebraic definition of a similar
operation characterising the relation~$\in^\sel$.
Instead, we have to define it directly in terms of~$\in^\sel$.
\begin{defi}
We define the \emph{selection operation}
$\sel : \bbT^\times\bbU\bbX \Rightarrow \bbU\bbX\bbT^\circ$ by
\begin{align*}
  \sel(t) := \set{ \langle\sigma,s\rangle }{ \varphi,\sigma : s \in^\sel t }\,.
\end{align*}
\upqed
\markenddef
\end{defi}

The properties of this operation are as follows.
\begin{lem}\label{Lem: basic properties of sel}\leavevmode
\begin{enuma}
\item $\sel : \bbT^\times\bbU\bbX \Rightarrow \bbU\bbX\bbT^\circ$ is a natural transformation
  on~$\Pos^{\Xi_+}$.
\item $\sel \circ \bbT^\times\pt = \pt \circ \un^+$
\item $\sel \circ \sing^\times = \bbU\bbX\sing$
\item $\sel \circ \bbT^\times(\pt \circ \IN) = \pt \circ \un$
\item $\bbU(\dun \circ \re_0) \circ \sel \circ \bbT^\times\bbU\un =
         \sel \circ \bbT^\times\bbU\un$
\end{enuma}
\end{lem}
\begin{proof}
(a)
Let $f : A \to B$. Then
\begin{alignat*}{-1}
  &\varphi,\sigma : s \in^\sel \bbT^\times\bbU\bbX f(t) \\
  \iff\quad
  &\langle\varphi_{/v},s(v)\rangle \in \bbU\bbX f(t(\varphi(v)))\,,
    &&\quad\text{for all } v\,, \\
  \iff\quad
  &s(v) \geq f(r(v)) \qtextq{and} \langle\varphi_{/v},r(v)\rangle \in t(\varphi(v))\,,
   &&\quad\text{for all } v\,, \\
  \iff\quad
  &s \geq \bbT^\times f(r) \qtextq{and} \varphi,\sigma : r \in^\sel t\,,
\end{alignat*}
implies that $\sel(\bbT^\times\bbU\bbX f(t)) = \bbU\bbX\bbT^\times f(\sel(t))$.

(b)
To simplify notation, we will again leave the universal quantification over vertices~$v$
implicit.
Let $t \in \bbT^\times\bbX A$. Then
\begin{align*}
  \sel(\bbT^\times\pt(t))
  &= \Aboveseg\set{ \langle\sigma,s\rangle }{ \varphi,\sigma : s \in^\sel \bbT^\times\pt(t) } \\
  &= \begin{aligned}[t]
       \Aboveseg\biglset \langle\sigma,s\rangle \bigmset {}
         &\langle\varphi_{/v},s(v)\rangle \in \pt(t(\varphi(v))) \text{ or} \\
         &[s(v) = x \text{ and } \bbT^\times\pt(t)(\varphi(v)) = \sigma(x)] \bigrset
     \end{aligned} \displaybreak[0]\\
  &= \begin{aligned}[t]
       \Aboveseg\biglset \langle\sigma,s\rangle \bigmset {}
         &\langle\varphi_{/v},s(v)\rangle \geq t(\varphi(v)) \text{ or} \\
         &[s(v) = x \text{ and } t(\varphi(v)) = \sigma(x)] \bigrset
     \end{aligned} \displaybreak[0]\\
  &= \begin{aligned}[t]
       \Aboveseg\biglset \langle\sigma,s\rangle \bigmset {}
         &\langle\varphi_{/v},s(v)\rangle = t(\varphi(v)) \text{ or} \\
         &[s(v) = x \text{ and } t(\varphi(v)) = \sigma(x)] \bigrset
     \end{aligned} \displaybreak[0]\\
  &= \Aboveseg\set{ \langle\sigma,s\rangle }{ \varphi,\sigma : s =^\sel t } \\
  &= \Aboveseg\{ \un^+(t) \} \\
  &= \pt(\un^+(t))\,.
\end{align*}

(c)
Let $I \in \bbU\bbX A$. Then
\begin{align*}
  \sel(\sing^\times(I))
  &= \Aboveseg\set{ \langle\sigma,s\rangle }{ \varphi,\sigma : s \in^\sel \sing^\times(I) } \\
  &= \Aboveseg\set{ \langle\sigma,s\rangle }
       { s = \sing(a)\,,\ \langle\tau,a\rangle \in I\,,\ \sigma = \tau } \\
  &= \Aboveseg\set{ \langle\sigma,\sing(a)\rangle }{ \langle\sigma,a\rangle \in I } \\
  &= \bbU\bbX\sing(I)\,.
\end{align*}

(d) By~(b) and Lemma~\ref{Lem: dun commutes with Flat}\,(d), we have
\begin{align*}
  \sel \circ \bbT^\times(\pt \circ \IN)
  = \pi \circ \un^+ \circ \bbT^\times\IN
  = \pi \circ \un\,.
\end{align*}

(e)
Let $\langle\sigma,s\rangle \in \sel(\bbT^\times\bbU\un(t))$.
Then $\varphi,\sigma : s \in^\sel \bbT^\times\bbU\un(t)$, which implies that
\begin{align*}
  \langle\varphi_{/v},s(v)\rangle \in \un(t(\varphi(v)))\,.
\end{align*}
Consequently, we have $\un(s(v)) = \langle\id,s(v)\rangle$, that is,
$\bbT^\circ\un(s) = \bbT^\circ\IN(s)$.
Hence,
\begin{align*}
  (\bbT^\circ\un \circ \re_0)(\langle\sigma,s\rangle)
  &= \bbT^\circ\un({}^\sigma s) \\
  &= {}^\sigma\bbT^\circ\un(s) \\
  &= {}^\sigma\bbT^\circ\IN(s) \\
  &= \bbT^\circ\IN({}^\sigma s)
   = (\bbT^\circ\IN \circ \re_0)(\langle\sigma,s\rangle)\,.
\end{align*}
Furthermore, $s \in \bbT^\circ A$ implies that $\un(s) = \langle\id,s\rangle$.
It therefore follows by Lemma~\ref{Lem: dun commutes with Flat}\,(d) that
\begin{align*}
  (\dun \circ \re_0)(\langle\sigma,s\rangle)
  &= (\un^+ \circ \bbT^\times\un \circ \re_0)(\langle\sigma,s\rangle) \\
  &= (\un^+ \circ \bbT^\times\IN \circ \re_0)(\langle\sigma,s\rangle) \\
  &= (\un \circ \re_0)(\langle\sigma,s\rangle) \\
  &= (\comp \circ \bbX\un)(\langle\sigma,s\rangle) \\
  &= (\comp \circ \bbX\IN)(\langle\sigma,s\rangle) \\
  &= \langle\sigma,s\rangle\,.
\end{align*}
Consequently,
\begin{align*}
  \bbU(\dun \circ \re_0) \restriction (\sel \circ \bbT^\times\bbU\un) =
    \bbU\id \restriction (\sel \circ \bbT^\times\bbU\un)\,.
\end{align*}
\upqed
\end{proof}

We need one more equation concerning the operation~$\sel$ whose proof is more involved\?:
Lemma~\ref{Lem: flat in^un flat} below contains a commutation relation between $\sel$
and $\Flat^\times$ that is similar to one of the axioms of a distributive law.
The proof makes use of the following technical lemma.
\begin{lem}\label{Lem: successor maps are equal}
Let $r \in \bbT^\times\bbT^\times A$ and $t \in \bbT^\times\bbT^\times B$ be trees,
set $s := \Flat(r)$, let
\begin{align*}
  \chi    &: \dom(s) \to \dom(\Flat^\times(t))\,, \\
  \varphi &: \dom(r) \to \dom(t)\,, \\
  \psi_v  &: \dom(r(v)) \to \dom(t(\varphi(v)))
\end{align*}
be surjective graph homomorphisms, and let
\begin{align*}
  \lambda &: \dom(\Flat^\times(t)) \to
              \sum_{v \in \dom_0(t)} \dom_0(t(v)) + [\dom(t) \setminus \dom_0(t)]\,, \\
  \mu &: \dom(\Flat(r)) \to
          \sum_{v \in \dom_0(r)} \dom_0(r(v)) + [\dom(r) \setminus \dom_0(r)]
\end{align*}
be the functions induced by the canonical maps
\begin{align*}
  \dom_0(\Flat^\times(t)) &\to \sum_{v \in \dom_0(t)} \dom_0(t(v)) \\
  \dom_0(\Flat(r)) &\to \sum_{v \in \dom_0(r)} \dom_0(r(v))\,.
\end{align*}
Then
\begin{align*}
  \lambda(\chi(w)) = \langle\varphi(v),\psi_v(u)\rangle\,,
  \quad\text{for every } w \in \dom_0(s) \text{ with } \mu(w) = \langle v,u\rangle\,,
\end{align*}
implies that
\begin{align*}
  \chi_{/w} = (\psi_v)_{/u}\,,
  \quad\text{for } \mu(w) = \langle v,u\rangle\,.
\end{align*}
\end{lem}
\begin{proof}
Consider a vertex $w \in \dom_0(s)$ with $\mu(w) = \langle v,u\rangle$
and an $x$-successor~$\tilde u$ of~$u$.
Suppose that $\lambda(\chi(w)) = \langle v',u'\rangle$.
First, let us consider the case where $\tilde u \in \dom_0(r(v))$.
Let $\tilde w$~be the successor of~$w$ with $\mu(\tilde w) = \langle v,\tilde u\rangle$.
By assumption, we have $\lambda(\chi(\tilde w)) = \langle\varphi(v),\psi_v(\tilde u)\rangle$
and $\psi_v(\tilde u)$ is the $y$-successor of~$\psi_v(u)$ in~$t(\varphi(v))$, for some~$y$.
By definition, it follows that $\chi_{/w}(x) = y$ and $(\psi_v)_{/u}(x) = y$.

It remains to consider the case where $\tilde u \notin \dom_0(r(v))$.
Then $r(v)(\tilde u) = z$, for some variable~$z$.
Let $v'$~be the $z$-successor of~$v$, let $\emptyseq$~be the root of~$r(v')$,
and let $\tilde w$~be the successor of~$w$ with $\mu(\tilde w) = \langle v',\emptyseq\rangle$.
Then $\lambda(\chi(w)) = \langle\varphi(v'),\psi_{v'}(\emptyseq)\rangle$.
Let $y$~be the variable such that $\lambda(\varphi(v'),\psi_{v'}(\emptyseq))$ is the
$y$-successor of $\lambda(\varphi(v),\psi_v(u))$.
Then $\chi_{/w}(x) = y$ and $(\psi_v)_{/u}(x) = y$.
\end{proof}

\begin{lem}\label{Lem: flat in^un flat}
$\sel \circ \Flat^\times = \bbU\bbX\Flat \circ \sel \circ \bbT^\times\sel$
\end{lem}
\begin{proof}
Note that the canonical function
\begin{align*}
  \dom_0(\Flat^\times(t)) \to \sum_{v \in \dom_0(t)} \dom_0(t(v))
\end{align*}
induces a function
\begin{align*}
  \lambda : \dom(\Flat^\times(t)) \to
              \sum_{v \in \dom_0(t)} \dom_0(t(v)) + [\dom(t) \setminus \dom_0(t)]\,.
\end{align*}
Similarly, for a tree~$r$ (which we will specify below), we obtain a function
\begin{align*}
  \mu : \dom(\Flat(r)) \to
          \sum_{v \in \dom_0(r)} \dom_0(r(v)) + [\dom(r) \setminus \dom_0(r)]\,.
\end{align*}

To prove the lemma, we check the two inclusions separately.

$(\supseteq)$
Suppose that $\langle\sigma,s\rangle \in \bbU\bbX\Flat(\sel(\bbT^\times\sel(t)))$.
Then
\begin{align*}
  s = \Flat(r)
  \qtextq{for some }
  \varphi,\sigma : r \in^\sel \bbT^\times\sel(t)\,.
\end{align*}
For every vertex~$v$ of~$r$, it follows that
\begin{align*}
  \langle\varphi_{/v},r(v)\rangle \in \sel(t(\varphi(v)))
  \qtextq{or}
  r(v) = x \text{ and } \sel(t(\varphi(v))) = \sigma(x)\,.
\end{align*}
This implies that
\begin{align*}
  \psi_v,\varphi_{/v} : r(v) \in^\sel t(\varphi(v))
  \qtextq{or}
  r(v) = x \text{ and } t(\varphi(v)) = \sigma(x)\,,
\end{align*}
for some homomorphism~$\psi_v$.
Let $\chi$~be the unique graph homomorphism satisfying the equations
\begin{align*}
  \lambda(\chi(w)) =
    \begin{cases}
      \langle\varphi(v),\psi_v(u)\rangle &\text{if } \mu(w) = \langle v,u\rangle\,, \\
      \varphi(v)                         &\text{if } \mu(w) = v\,,
    \end{cases}
\end{align*}
where $\lambda$~and~$\mu$ are the homomorphisms defined above.
We claim that $\chi,\sigma : s \in^\sel \Flat^\times(t)$,
which implies that $\langle\sigma,s\rangle \in \sel(\Flat^\times(t))$.

Hence, fix a vertex $w \in \dom(s) = \dom(\Flat(r))$.
First, consider the case where $w \in \dom_0(s)$.
Suppose that $\mu(w) = \langle v,u\rangle$.
Then $\psi_v,\varphi_{/v} : r(v) \in^\sel t(\varphi(v))$ implies that
\begin{align*}
  \langle(\psi_v)_{/u},r(v)(u)\rangle \in t(\varphi(v))(\psi_v(u))\,.
\end{align*}
Consequently, we have
\begin{align*}
  \langle(\psi_v)_{/u},s(w)\rangle \in t(\varphi(v))(\psi_v(u)) = \Flat^\times(t)(\chi(w))\,.
\end{align*}
Furthermore, we have $(\psi_v)_{/u} = \chi_{/w}$ by Lemma~\ref{Lem: successor maps are equal}.

It remains to consider the case where $s(w) = x$ is a variable.
Then $\mu(w) = v$, for some $v \in \dom(r)$, and $r(v) = x$ implies
that $t(\varphi(v)) = \sigma(x)$.
Hence,
\begin{align*}
  \Flat^\times(t)(\chi(w)) = t(\lambda(\chi(w))) = t(\varphi(v)) = \sigma(x)\,.
\end{align*}

$(\subseteq)$
Suppose that $\langle\sigma,s\rangle \in \sel(\Flat^\times(t))$.
Then
\begin{align*}
  \chi,\sigma : s \in^\sel \Flat^\times(t)\,,
  \quad\text{for some } \chi\,.
\end{align*}
We define a tree~$r$ with $\Flat(r) = s$ as follows.
Intuitively, we factorise~$s$ by cutting every edge $w \to w'$
such that the corresponding vertices $\chi(w)$ and $\chi(w')$ in~$\Flat^\times(t)$
belong to different components $t(v)$ and $t(v')$, i.e., if
$\lambda(\chi(w)) = \langle v,u\rangle$ and $\lambda(\chi(w')) = \langle v',u'\rangle$ with
$v \neq v'$.
The formal definition is as follows.
Let us call a vertex $w \in \dom(s)$ \emph{principal} if its image under~$\chi$
corresponds to the root of some conponent~$t(v)$, or to a leaf, that is, if
\begin{align*}
  \lambda(\chi(w)) = \langle v,\emptyseq\rangle \qtextq{or}
  \lambda(\chi(w)) = v\,, \quad\text{for some } v \,,
\end{align*}
(where $\emptyseq$~denotes the root of~$t(v)$).
We define the domain of~$r$ by
\begin{align*}
  \dom(r) := \set{ w \in \dom(s) }{ w \text{ is principal} }
\end{align*}
and the edge relation as follows.
Given a principal vertex~$w$, let $w_0,\dots,w_{n-1}$ be an enumeration of all
minimal principal vertices~$w'$ with $w \prec w'$.
We make~$w_i$ an $i$-successor of~$w$. (The precise labels~$i$ are not
important, only the fact that they are pairwise distinct.)
Finally, the labelling of~$r$ is given by
\begin{align*}
  r(w) := \begin{cases}
            r_w  &\text{if } w \in \dom_0(s)\,, \\
            s(w) &\text{if } w \notin \dom_0(s)\,,
          \end{cases}
\end{align*}
where $r_w$~is the tree with
\begin{align*}
  \dom(r_w) &:=
    \begin{aligned}[t]
      \biglset u \in \dom(s) \bigmset {}
        & w \preceq u \text{ and there is no principal } w' \text{ with} \\
        & w \prec w' \prec u \bigrset\,,
    \end{aligned} \\
  r_w(u) &:= \begin{cases}
              s(u) &\text{if } u \notin \dom(r) \text{ or } u = w\,, \\
              i    &\text{if } u = w_i \in \dom(r) \text{ is the $i$-successor of } w
                    \text{ in } r\,.
            \end{cases}
\end{align*}

By definition, it follows that $\Flat(r) = s$ and that
\begin{alignat*}{-1}
  \mu(w) &= \langle v,w\rangle\,,
  &&\quad\text{if } w \in \dom_0(s)\,,
  \text{ where } v \text{ is the maximal principal} \\
  &&&\qquad\text{vertex with } v \preceq w\,, \\
\text{and~~}
  \mu(w) &= w\,,
  &&\quad\text{if } w \notin \dom_0(s)\,.
\end{alignat*}
Let $\varphi$~and~$\psi_v$ be the functions defined by the equations
\begin{alignat*}{-1}
  \langle\varphi(v),\psi_v(u)\rangle &= \lambda(\chi(w))\,,
    &&\quad\text{for } \mu(w) = \langle v,u\rangle\,, \\
  \varphi(w) &= \lambda(\chi(w))\,, &&\quad\text{if } w \in \dom(s) \setminus \dom_0(s)\,, \\
  \psi_v(u)  &= u''                 &&\quad\text{if } u \in \dom(r(v)) \setminus \dom_0(r(v))\,,
\end{alignat*}
where the vertex~$u''$ in the last equation is chosen as follows.
Given~$u$, let $u'$~be the predecessor of~$u$ and
let $x$~be the label of the edge $u' \to u$.
Then $u''$~is the $(\psi_v)_{/u'}(x)$-successor of~$\psi_v(u')$.

We claim that, for all~$v$,
\begin{align*}
  \psi_v,\varphi_{/v} : r(v) \in^\sel t(\varphi(v))
  \qtextq{or}
  r(v) = x \text{ and } t(\varphi(v)) = \sigma(x)\,.
\end{align*}
Then it follows that
\begin{align*}
  \langle\varphi_{/v},r(v)\rangle \in \sel(t(\varphi(v)))
  \qtextq{or}
  r(v) = x \text{ and } \sel(t(\varphi(v))) = \sigma(x)\,.
\end{align*}
Thus,
\begin{align*}
  \langle\sigma,r\rangle \in \sel(\bbT^\times\sel(t))
  \qtextq{and}
  \langle\sigma,s\rangle \in \bbU\bbX\Flat(\sel(\bbT^\times\sel(t)))\,,
\end{align*}
as desired.
Hence, it remains to prove the above claim.

If $r(v) = x$ is a variable, we have $s(v) = r(v) = x$ and, therefore,
\begin{align*}
  t(\varphi(v)) = t(\lambda(\chi(v))) = \Flat^\times(t)(\chi(v)) = \sigma(x)\,,
\end{align*}
as desired.
Otherwise, $v \in \dom_0(r)$ and we have to show that
\begin{align*}
  \psi_v,\varphi_{/v} : r(v) \in^\sel t(\varphi(v))\,.
\end{align*}
Note that $\chi,\sigma : s \in^\sel \Flat^\times(t)$ implies that
\begin{align*}
  \langle\chi_{/w},s(w)\rangle \in \Flat^\times(t)(\chi(w))\,,
  \quad\text{for all } w\,.
\end{align*}
We distinguish two cases.
If $u \in \dom_0(r(v))$, let $w \in \dom_0(s)$ be the vertex with
$\mu(w) = \langle v,u\rangle$. Then
\begin{align*}
  \langle\chi_{/w},r(v)(u)\rangle = \langle\chi_{/w},s(w)\rangle
  \in \Flat^\times(t)(\chi(w)) = t(\varphi(v))(\psi_v(u))\,.
\end{align*}
By Lemma~\ref{Lem: successor maps are equal}, we have $\chi_{/w} = (\psi_v)_{/u}$,
which implies that
\begin{align*}
  \langle(\psi_v)_{/u},r(v)(u)\rangle \in t(\varphi(v))(\psi_v(u))\,.
\end{align*}

If $u \in \dom(r(v)) \setminus \dom_0(r(v))$ with label $r(v)(u) = x$,
let $v'$~be the $x$-successor of~$v$.
By definition of~$\varphi_{/v}$,
it follows that $\varphi(v')$~is the $\varphi_{/v}(x)$-successor of~$\varphi(v)$ in~$t$.
This implies that $t(v)(\psi_v(u)) = \varphi_{/v}(x)$.
\end{proof}

\subsection{A partial distributive law}   

The idea to find our partial distributive law is to work in the category
of unravelling structures, although this does not solve our problems entirely.
First of all, there is no obvious way to lift the functor~$\bbU$ to unravelling structures.
Given an unravelling structure~$A$, we can define an `unravelling map'
$\bbU\un : \bbU A \to \bbU\bbX A$, but we would need one of the form $\bbU A \to \bbX\bbU A$,
and there is no natural transformation $\bbU\bbX \Rightarrow \bbX\bbU$.
The functor~$\bbT^\times$ on the other hand \emph{can} be lifted to the category
of unravelling structures, but only in a trivial way\?:
given~$A$ we can forget its unravelling structure, construct~$\bbT^\times A$,
and equip it with the canonical unravelling structure
defined above (which does not depend on that of~$A$).
In particular, with this definition the monad multiplication $\Flat^\times$ would
not be a morphism of the resulting unravelling structure.
What would be more useful would be a lift that uses deep unravelling~$\dun$ as the unravelling
operation on~$\bbT^\times A$. But there is no corresponding reconstitution operation~$\re$
satisfying $\re \circ \dun = \id$.

What we will do instead is to use an ad-hoc argument showing how to define a lift of~$\bbU$
to sufficiently well-behaved $\bbT^\times$-algebras.
We are mainly interested in free $\bbT^\times$-algebras, but a slightly more abstract
definition helps to make the proof more modular.
We extract the needed properties of the algebras in question in the following technical
definition.
\begin{defi}
We say that a $\bbT^\times$-algebra $\frakA = \langle A,\pi\rangle$ \emph{supports unravelling}
if its universe~$A$ can be equipped with an unravelling structure that
satisfies the following conditions.
\begin{align*}
  \pi \circ \re \circ \bbX\sing^\times &= \re\,, \\
  \un \circ \re &= \comp \circ \bbX\un\,, \\
  \bbX(\un \circ \pi \circ \iota) \circ \dun
    &= \bbX(\IN \circ \pi \circ \iota) \circ \dun\,.
\end{align*}
\upqed
\markenddef
\end{defi}

The intended target for this definition are the free algebras.
We start by noting that these satisfy the above conditions.
\begin{prop}
The free $\bbT^\times$-algebra $\langle \bbT^\times A,\Flat^\times\rangle$ supports unravelling.
\end{prop}
\begin{proof}
Using the operations $\un$~and~$\re$ from Definitions \ref{Def: un}~and~\ref{Def: re},
it follows by Lemma~\ref{Lem: un and re for trees} (e)~and~(b), that
\begin{align*}
  \Flat^\times \circ \re_0 \circ \bbX\sing
  &= \re_0 \circ \bbX(\Flat^\times \circ \iota) \circ \bbX\sing \\
  &= \re_0 \circ \bbX(\Flat^\times \circ \sing^\times)
   = \re_0\,, \\
  \un \circ \re_0
  &= \comp \circ \bbX\un\,,
\end{align*}
while the third condition follows by Lemma~\ref{Lem: dun commutes with Flat}\,(a).
\end{proof}

For the proof below,
let us collect a few basic properties of algebras that support unravelling.
\begin{lem}\label{Lem: un(pi(s))}
Let\/ $\frakA$~be a $\bbT^\times$-algebra that supports unravelling.
\begin{enuma}
\item $\pi \circ \re = \re \circ \bbX\pi$
\item $\pi \circ \re \circ \dun = \pi$
\item $\un \circ \pi \circ \re \circ \dun = \bbX\pi \circ \dun$
\item $\un \circ \pi = \bbX\pi \circ \dun$
\item $\bbU(\un \circ \pi \circ \re_0) \circ \sel \circ \bbT^\times\bbU\un =
         \bbU\bbX\pi \circ \sel \circ \bbT^\times\bbU\un$
\end{enuma}
\end{lem}
\begin{proof}
Below we will make freely use of the equations from Lemma~\ref{Lem: un and re for trees}.

(a) We have
\begin{align*}
  \pi \circ \re
  &= \pi \circ \re \circ \bbX(\Flat^\times \circ \sing^\times) \\
  &= \pi \circ \Flat^\times \circ \re \circ \bbX\sing^\times \\
  &= \pi \circ \bbT^\times\pi \circ \re \circ \bbX\sing^\times \\
  &= \pi \circ \re \circ \bbX\bbT^\circ\pi \circ \bbX\sing^\times \\
  &= \pi \circ \re \circ \bbX\sing^\times \circ \bbX\pi \\
  &= \re \circ \bbX\pi\,,
\end{align*}
where the last step follows from the fact that $\frakA$~supports unravelling.

(b)
Since
\begin{align*}
  \pi \circ \re_0 \circ \bbX\sing \circ \un
  = \re_0 \circ \bbX\pi \circ \bbX\sing \circ \un
  = \re_0 \circ \un
  = \id\,,
\end{align*}
we have
\begin{align*}
  \pi \circ \re \circ \dun
  &= \pi \circ \re \circ \un \circ \Flat^\times \circ
        \bbT^\times(\re_0 \circ \bbX\sing \circ \un) \\
  &= \pi \circ \Flat^\times \circ \bbT^\times(\re_0 \circ \bbX\sing \circ \un) \\
  &= \pi \circ \bbT^\times\pi \circ \bbT^\times(\re_0 \circ \bbX\sing \circ \un) \\
  &= \pi \circ \bbT^\times\id \\
  &= \pi\,.
\end{align*}

(c)
By~(a) and the fact that $\frakA$~supports unravelling, we have
\begin{align*}
  \un \circ \pi \circ \re \circ \dun
  &= \un \circ \re \circ \bbX\pi \circ \dun \\
  &= \comp \circ \bbX\un \circ \bbX\pi \circ \dun \\
  &= \comp \circ \bbX(\IN \circ \pi) \circ \dun
   = \bbX\pi \circ \dun\,.
\end{align*}

(d) By (c)~and~(b), we have
\begin{align*}
  \bbX\pi \circ \dun
   = \un \circ \pi \circ \re \circ \dun
   = \un \circ \pi\,.
\end{align*}

(e)
By~(a), Lemma~\ref{Lem: basic properties of sel}\,(e), and the fact that
$\frakA$~supports unravelling, we have
\begin{align*}
     & \bbU(\un \circ \pi \circ \re) \circ \sel \circ \bbT^\times\bbU\un \\
{}={}& \bbU(\un \circ \re \circ \bbX\pi) \circ \sel \circ \bbT^\times\bbU\un \\
{}={}& \bbU(\comp \circ \bbX\un \circ \bbX\pi) \circ \sel \circ \bbT^\times\bbU\un \\
{}={}& \bbU(\comp \circ \bbX(\un \circ \pi) \circ \dun \circ \re_0) \circ \sel
         \circ \bbT^\times\bbU\un \\
{}={}& \bbU(\comp \circ \bbX(\IN \circ \pi) \circ \dun \circ \re_0) \circ \sel
         \circ \bbT^\times\bbU\un \\
{}={}& \bbU(\comp \circ \bbX(\IN \circ \pi)) \circ \sel \circ \bbT^\times\bbU\un \\
{}={}& \bbU\bbX\pi \circ \sel \circ \bbT^\times\bbU\un\,.
\end{align*}
\upqed
\end{proof}

Finally we can state our partial distributive law for $\bbU$~and~$\bbT^\times$
for algebras that support unravelling.
\begin{prop}\label{Prop: lift to algebras supporting unravelling}
If\/ $\frakA = \langle A,\pi\rangle$ is a\/ $\bbT^\times$-algebra supporting unravelling,
we can form a\/ $\bbT^\times$-algebra\/ $\bbU\frakA := \langle\bbU A,\hat\pi\rangle$ with product
\begin{align*}
  \hat\pi := \bbU(\pi \circ \re_0) \circ \sel \circ \bbT^\times\bbU\un\,.
\end{align*}
Furthermore, the function\/ $\pt : A \to \bbU A$ induces an embedding\/
$\frakA \to \bbU\frakA$.
\end{prop}
\begin{proof}
We have to check three equations.
To see that $\pt$~is an embedding, note that
\begin{align*}
  \hat\pi \circ \bbT^\times\pt
  &= \bbU(\pi \circ \re_0) \circ \sel \circ \bbT^\times\bbU\un \circ \bbT^\times\pt \\
  &= \bbU(\pi \circ \re_0) \circ \sel \circ \bbT^\times\pt \circ \bbT^\times\un \\
  &= \bbU(\pi \circ \re_0) \circ \pt \circ \un^+ \circ \bbT^\times\un \\
  &= \pt \circ \pi \circ \re_0 \circ \un^+ \circ \bbT^\times\un \\
  &= \pt \circ \pi \circ \re_0 \circ \dun \\
  &= \pt \circ \pi \\
  &= \bbU\pi \circ \pt\,.
\end{align*}
where the third step follows by Lemma~\ref{Lem: basic properties of sel}\,(b)
and the sixth one by Lemma~\ref{Lem: un(pi(s))}\,(b).
For the unit law, we have
\begin{align*}
  \hat\pi \circ \sing^\times
  &= \bbU(\pi \circ \re_0) \circ \sel \circ \bbT^\times\bbU\un \circ \sing^\times \\
  &= \bbU(\pi \circ \re_0) \circ \sel \circ \sing^\times \circ \bbU\un \\
  &= \bbU(\pi \circ \re_0) \circ \bbU\bbX\sing \circ \bbU\un \\
  &= \bbU(\pi \circ \sing \circ \re_0 \circ \un) \\
  &= \bbU(\id \circ \id) \\
  &= \id\,,
\end{align*}
where the third step follows by Lemma~\ref{Lem: basic properties of sel}\,(c).
Finally, for the associative law,
\begin{align*}
  \hat\pi \circ \bbT^\times\hat\pi
  &= \bbU(\pi \circ \re_0) \circ \sel \circ \bbT^\times\bbU\un \circ
       \bbT^\times(\bbU(\pi \circ \re_0) \circ \sel \circ \bbT^\times\bbU\un) \\
  &= \bbU(\pi \circ \re_0) \circ \sel \circ
       \bbT^\times(\bbU(\un \circ \pi \circ \re_0) \circ \sel \circ \bbT^\times\bbU\un) \\
  &= \bbU(\pi \circ \re_0) \circ \sel \circ
       \bbT^\times(\bbU\bbX\pi \circ \sel \circ \bbT^\times\bbU\un) \\
  &= \bbU(\pi \circ \re_0) \circ \bbU\bbX\bbT^\circ\pi \circ \sel \circ
       \bbT^\times(\sel \circ \bbT^\times\bbU\un) \\
  &= \bbU(\pi \circ \bbT^\circ\pi \circ \re_0) \circ \sel \circ
       \bbT^\times(\sel \circ \bbT^\times\bbU\un) \\
  &= \bbU(\pi \circ \Flat^\times \circ \re_0) \circ \sel \circ
       \bbT^\times(\sel \circ \bbT^\times\bbU\un) \\
  &= \bbU(\pi \circ \re_0 \circ \bbX\Flat^\times) \circ \sel \circ
       \bbT^\times(\sel \circ \bbT^\times\bbU\un) \\
  &= \bbU(\pi \circ \re_0) \circ \sel \circ \Flat^\times \circ
       \bbT^\times\bbT^\times\bbU\un \\
  &= \bbU(\pi \circ \re_0) \circ \sel \circ \bbT^\times\bbU\un \circ \Flat^\times \\
  &= \hat\pi \circ \Flat^\times\,.
\end{align*}
where the third step follows by Lemma~\ref{Lem: un(pi(s))}\,(e)
and the eighth one by Lemma~\ref{Lem: flat in^un flat}.
\end{proof}

For technical reasons, we have worked so far in the category $\Pos^{\Xi_+}$.
But the category we are actually interested in is~$\Pos^\Xi$.
The following consequence can be considered the main result of this section.
\begin{thm}\label{Thm: UTA is T-algebra}
In\/ $\Pos^\Xi$, the set\/~$\bbU\bbT^\times A$ forms a\/ $\bbT^\times$-algebra with product
\begin{align*}
  \hat\pi(t) := \Aboveseg\set{ \Flat^\times({}^\sigma s) }{ \varphi,\sigma : s \in^\un t }\,.
\end{align*}
\end{thm}
\begin{proof}
We know by Proposition~\ref{Prop: lift to algebras supporting unravelling} that
$\bbU\bbT^\times A^\uparrow$ forms a $\bbT^\times$-algebra in~$\Pos^{\Xi_+}$.
Since $\bbU\bbT^\times A = (\bbU\bbT^\times A^\uparrow)|_\Xi$, the claim follows
by Lemma~\ref{Lem: reduct is an algebra}.
\end{proof}

In order to strengthen this theorem to obtain a $\bbU\bbT^\times$-algebra,
we would need to prove that $\bbU\bbT^\times$~forms a monad.
The next result shows that the canonical choice for the corresponding monad multiplication
does not work.
(Note that this is not a simple consequence of Theorem~\ref{Thm: dist iff linear}
since it might be the case that, instead of condition~\textsc{(m1)} of
Theorem~\ref{Thm: distributive law of monads}\,(4), it is \textsc{(m2)}~or~\textsc{(m3)}
that is violated.)
\begin{prop}
The function $\kappa : \bbU\bbT^\times\bbU\bbT^\times A \to \bbU\bbT^\times A$ with
\begin{align*}
  \kappa(T) :=
    \Aboveseg\set{ \Flat^\times({}^\sigma s) }{ \varphi,\sigma : s \in^\un t\,,\ t \in T }
\end{align*}
does \emph{not} satisfy the associative law
\begin{align*}
  \kappa \circ \kappa = \kappa \circ \bbU\bbT^\times\kappa\,.
\end{align*}
\end{prop}
\begin{proof}
We use term notation $a(c),b(c,d),\dots$ for trees.
Note that, for two sets
\begin{align*}
  X = \set{ a_i(x_0,x_0) }{ i < m }
  \qtextq{and}
  Y = \set{ \sing^\times(c_i) }{ i < n }
\end{align*}
(where $a_i \in A_2$ and $c_i \in A_0$) we have
\begin{align*}
  \kappa(\{X(Y)\})
  &= \set{ \Flat^\times({}^\sigma s) }{ \varphi,\sigma : s \in^\un X(Y) } \\
  &= \begin{aligned}[t]
       \biglset \Flat^\times({}^\sigma s) \bigmset {}
       &s = u(v,w)\,,\ u = \sing^\times(a_i)\,, \\
       &v = \sing^\times(c_k)\,,\ w = \sing^\times(c_l)\,,\ i < m\,,\ k,l < n \bigrset
     \end{aligned}\\
  &= \set{ a_i(c_k,c_l) }{ i < m\,,\ k,l < n }\,.
\end{align*}
Similarly, if the $a_i \in A_1$ are unary, we obtain
\begin{align*}
  \kappa(\{X(Y)\}) = \set{ a_i(c_k) }{ i < m\,,\ k < n }\,.
\end{align*}
Setting
\begin{alignat*}{-1}
  I &:= \{a(x_0,x_0)\}\,,
  &\qquad
  C &:= \{c\}\,, \\
  J &:= \{b(x_0,x_0)\}\,,
  &\qquad
  D &:= \{d\}\,, \\
  K &:= \{\sing^\times(I),\sing^\times(J)\}\,,
  &\qquad
  E &:= \{\sing^\times(C),\sing^\times(D)\}\,,
\end{alignat*}
we obtain
\begin{align*}
  \kappa(\{K(E)\}) &= \{I(C),I(D),J(C),J(D)\}\,, \\[1ex]
  \kappa(\{I(C)\}) &= \{a(c,c)\}\,, \qquad
  \kappa(\{I(D)\})  = \{a(d,d)\}\,, \\
  \kappa(\{J(C)\}) &= \{b(c,c)\}\,, \qquad
  \kappa(\{J(D)\})  = \{b(d,d)\}\,, \\[1ex]
  (\kappa\circ\kappa)(\{K(E)\}) &= \{a(c,c),a(d,d),b(c,c),b(d,d)\}\,, \\[1ex]
  \kappa(K) &= \rlap{I \cup J}\hphantom{C \cup D}{} =: X\,, \\
  \kappa(E) &= C \cup D =: Y\,, \\[1ex]
  \bbU\bbT^\times\kappa(\{K(E)\}) &= \{ X(Y) \}\,, \\[1ex]
  (\kappa\circ\bbU\bbT^\times\kappa)(\{K(E)\}) &=
    \set{ u(v,w) }{ u \in \{a,b\}\,,\ v,w \in \{c,d\} }\,.
\end{align*}
Hence,
\begin{align*}
  (\kappa\circ\kappa)(\{K(D)\}) \neq (\kappa\circ\bbU\bbT^\times\kappa)(\{K(D)\})\,.
\end{align*}
(For instance, the tree $a(c,d)$ does belong to the right-hand side, but not to the
left-hand one.)
\end{proof}

\section{Substitutions}   
\label{Sect:substitutions}

As a first application of the tools we have developed above, let us take a look at substitutions
for tree languages.
We present a simplified account of a recent result by Camino~et\,al.~\cite{CaminoDiDuMaSe22}
about finding solutions to inequalities of the form $\sigma[L] \subseteq R$
for regular tree languages $L$~and~$R$.
This simplification stems mainly from the terminology and notation introduced above.
It does not rely on the results we have proved, except for Lemma~\ref{Lem: sigma_oi is morphism},
which depends on Theorem~\ref{Thm: UTA is T-algebra}.
In the next section we will give a~second, more involved application that makes use of
Theorem~\ref{Thm: UTA is T-algebra} in a more substantial way.
\begin{defi}
Let $\Sigma$~be an alphabet.

(a)
A~\emph{substitution} is a function $\sigma : X \to \bbU\bbT^\times\Sigma$.
We call~$\sigma$ \emph{regular} if every $\sigma(x) \subseteq \bbT^\times\Sigma$ is
a regular tree language.

(b)
A substitution $\sigma : X \to \bbU\bbT^\times\Sigma$
induces a function $\bbT^\times X \to \bbU\bbT^\times\Sigma$ in two different ways.
The \emph{inside-out} morphism~$\sigma_{\mathrm{io}}$ is defined by
\begin{align*}
  \sigma_{\mathrm{io}}(t) := \set{ \Flat^\times(s) }{ s \in^\bbR \bbR\sigma(t) }\,,
\end{align*}
while the \emph{outside-in} morphism~$\sigma_{\mathrm{oi}}$ is defined by
\begin{align*}
  \sigma_{\mathrm{oi}}(t) :=
    \set{ \Flat^\times({}^\sigma s) }{ \varphi,\sigma : s \in^\un \bbR\sigma(t) }\,.
\end{align*}
\upqed
\markenddef
\end{defi}
\begin{rem}
(a) The reader should compare the simple definition above with the much more involved one
given in~\cite{CaminoDiDuMaSe22}.
As it turns out such simplifications are not uncommon when using the monadic framework.

(b)
Intuitively, the difference between these two variants is that, with the inside-out
version~$\sigma_{\mathrm{io}}$, we have to choose the same image $s(u) \in \sigma(t(v))$
for every vertex~$u$ of~$s$ corresponding to $v \in \dom(t)$, while the
outside-in~$\sigma_{\mathrm{oi}}$ version allows us to choose a different tree for each of them.
The former has the advantage of simplicity, but the latter turns out to be
more natural from an algebraic perspective\?: we will show below that it forms a morphism
of $\bbT^\times$-algebras.

(c)
In the notation of Section~\ref{Sect:non-linear trees}, we can rewrite the above definitions as
\begin{align*}
  \sigma_{\mathrm{io}} &= \bbU\Flat^\times \circ \dist \circ \bbT^\times\sigma\,, \\
  \sigma_{\mathrm{oi}} &=
    \bbU(\Flat^\times \circ \re_0) \circ \sel \circ \bbT^\times(\bbU\un \circ \sigma)\,.
\end{align*}
Hence, $\sigma_{\mathrm{io}}$~is based on the failed distributive law~$\dist$, while
$\sigma_{\mathrm{oi}}$~is based on the more successful attempt using the relation~$\in^\un$.
\markenddef
\end{rem}

For the next lemma, let us recall from Theorem~\ref{Thm: UTA is T-algebra} that
$\bbU\bbT^\times\Sigma$ indeed forms a $\bbT^\times$-algebra.
\begin{lem}\label{Lem: sigma_oi is morphism}
$\sigma_{\mathrm{oi}} : \bbT^\times X \to \bbU\bbT^\times\Sigma$ is a morphism of\/
$\bbT^\times$-algebras.
\end{lem}
\begin{proof}
According to Theorem~\ref{Thm: UTA is T-algebra},
the product of the algebra~$\bbU\bbT^\times\Sigma$ is given by
\begin{align*}
  \hat\pi := \bbU(\Flat^\times \circ \re_0) \circ \sel \circ \bbT^\times\bbU\un\,.
\end{align*}
Hence, $\sigma_{\mathrm{oi}} = \hat\pi \circ \bbT^\times\sigma$ and it follows that
\begin{align*}
  \sigma_{\mathrm{oi}} \circ \Flat^\times
  &= \hat\pi \circ \bbT^\times\sigma \circ \Flat^\times \\
  &= \hat\pi \circ \Flat^\times \circ \bbT^\times\bbT^\times\sigma \\
  &= \hat\pi \circ \bbT^\times\hat\pi \circ \bbT^\times\bbT^\times\sigma
   = \hat\pi \circ \bbT^\times\sigma_{\mathrm{oi}}\,.
\end{align*}
\upqed
\end{proof}
\begin{rem}
Note that the function $\sigma_{\mathrm{io}} : \bbT^\times X \to \bbU\bbT^\times\Sigma$
is \emph{not} a morphism of $\bbT^\times$-algebras.

\markenddef
\end{rem}

For the simpler inside-out substitutions, we can solve inequalities
$\rho_{\mathrm{io}}[L] \subseteq R$ as follows.
\begin{thmC}[\cite{CaminoDiDuMaSe22}]\label{Thm: matching problem}
Let $L \subseteq \bbT^\times X$ and $R \subseteq \bbT^\times\Sigma$ be regular tree
languages, $\sigma,\tau : X \to \bbU\bbT^\times\Sigma$ regular substitutions, and
let $S$~be the set of all substitutions~$\rho$ such that
\begin{align*}
  \sigma \subseteq \rho \subseteq \tau
  \qtextq{and}
  \rho_{\mathrm{io}}[L] \subseteq R\,.
\end{align*}
Then
\begin{enuma}
\item $S$~has finitely many maximal elements.
\item Every maximal element of~$S$ is regular.
\item We can effectively compute the maximal elements of~$S$.
\end{enuma}
\end{thmC}
\begin{proof}
Since $R$~is regular, it is recognised by some morphism $\eta : \bbT^\times\Sigma \to \frakA$
into a finitary $\bbT^\times$-algebra $\frakA = \langle A,\pi\rangle$
(for a proof see~\cite{Blumensath20,Blumensath21}).
We define the \emph{saturation} $\hat\rho : X \to \bbU\bbT^\times\Sigma$ of a given
substitution $\rho : X \to \bbU\bbT^\times\Sigma$ by
\begin{align*}
  \hat\rho(x) := \bigset{ s \in \bbT^\times\Sigma }{ \eta(s) \in \bbU\eta(\rho(x)) }\,.
\end{align*}
Then we have $\bbU\eta \circ \hat\rho = \bbU\eta \circ \rho$.
Note that we can rewrite the definition of~$\rho_{\mathrm{io}}$ as
\begin{align*}
  \rho_{\mathrm{io}} = \bbU\Flat^\times \circ \dist \circ \bbT^\times\rho\,.
\end{align*}
It follows that
\begin{align*}
  \bbU\eta \circ \rho_{\mathrm{io}}
  &= \bbU(\eta \circ \Flat^\times) \circ \dist \circ \bbT^\times\rho \\
  &= \bbU(\pi \circ \bbT^\times\eta) \circ \dist \circ \bbT^\times\rho
   = \bbU\pi \circ \dist \circ \bbT^\times\bbU\eta \circ \bbT^\times\rho\,.
\end{align*}
Consequently, we have
\begin{align*}
  \bbU\eta \circ \rho_{\mathrm{io}}
  &= \bbU\pi \circ \dist \circ \bbT^\times(\bbU\eta \circ \rho) \\
  &= \bbU\pi \circ \dist \circ \bbT^\times(\bbU\eta \circ \hat\rho)
   = \bbU\eta \circ \hat\rho_{\mathrm{io}}\,.
\end{align*}
As $\eta(s) = \eta(t)$ implies $s \in R \Leftrightarrow t \in R$,
it therefore follows that
\begin{align*}
  \rho_{\mathrm{io}}(t) \subseteq R \qtextq{implies} \hat\rho_{\mathrm{io}}(t) \subseteq R\,.
\end{align*}
Since $\rho \subseteq \hat\rho$ this implies that the maximal elements of~$S$ satisfy
$\rho = \hat\rho \cap \tau$. In particular, a substitution of this form is regular.
This proves~(b).

For~(a), note that the number of substitutions of the form~$\hat\rho$ is bounded by the number
of functions $X \to \bbU A$. As~$X$ is finite and $A$~is sort-wise finite,
there are only finitely many such functions.

It remains to establish~(c).
We can enumerate all functions $X \to \bbU A$.
This gives an enumeration of all substitutions of the form~$\hat\rho$.
For each of them, we can check whether $\sigma \leq \hat\rho \cap \tau$.
If so, $\hat\rho \cap \tau$ is a maximal element of~$S$.
Otherwise, it is not.
\end{proof}
The more complicated case of outside-in substitutions is still open.

\begin{rem}
There is one technical detail worth mentioning\?:
the way we have defined substitutions, every tree in~$\sigma(x)$, for $x \in X_\xi$,
contains \emph{all} variables in~$\xi$.
But usually one uses a more general notion of a substitution where the trees in~$\sigma(x)$
can omit some or all of these variables.
We can formalise this generalisation in our setting as follows.

We consider a substitution as a function $\sigma : X \to \bbU\bbT^<\Sigma$,
where $\bbT^<$~is the functor with
\begin{align*}
  \bbT^<_\xi X := \sum_{\zeta \subseteq \xi} \bbT^\times_\zeta X\,.
\end{align*}
We can extend the monad operation to~$\bbT^<$ in the obvious way.
As above we define two induced operations
$\sigma_{\mathrm{io}},\sigma_{\mathrm{oi}} : \bbT^< X \to \bbU\bbT^<\Sigma$.
The definition of the outside-in version is the same as above
\begin{align*}
  \sigma_{\mathrm{oi}}(t) :=
    \set{ \Flat^<({}^\sigma s) }{ \varphi,\sigma : s \in^\un \bbR^<\sigma(t) }
\end{align*}
(where $\bbR^<$~is the corresponding variant of~$\bbR$).

But the inside-out version is more complicated.
The problem is that some sets $\sigma(x)$~might be empty, but a tree~$t$
might still have a non-empty image $\sigma_{\mathrm{io}}(t)$ because,
for every vertex~$v$ with $\sigma(t(v)) = \emptyset$,
there might be some vertex~$u$ higher up in the tree
where we have chosen an element $s \in t(u)$ which omits the variable
corresponding to the subtree containing~$v$.
The easiest way to formalise this process is to make the problem disappear
by adding dummy elements to all sets~$\sigma(x)$.
Hence, fix some element $\bot \notin \Sigma$ and let
$\mu : \bbU\Sigma \to \bbU(\Sigma + \{\bot\})$
be the function with
\begin{align*}
  \mu(I) := I \cup \{\bot\}\,.
\end{align*}
Then we set
\begin{align*}
  \sigma_{\mathrm{io}}(t) :=
    \bigset{ \Flat^<(s) }{ s \in^{\bbR^<} \bbR^<(\mu \circ \sigma)(t)\,,\ 
                           \Flat^<(s) \in \bbT^<\Sigma }\,.
\end{align*}
The proof of Theorem~\ref{Thm: matching problem}
can now straightforwardly be adapted to these new definitions.
\markenddef
\end{rem}

\section{Regular expressions for infinite trees}   
\label{Sect:regex}

As a second, more involved application of our results let us define regular expressions
for languages of infinite trees.
Such expressions seem to be folklore, but we have not found them anywhere in the literature
(except for a few remarks in~\cite{Thomas90}).

We consider tree languages of the form $L \subseteq \bbT^\times_\xi\Sigma$,
for some alphabet~$\Sigma$ and some fixed sort $\xi \in \Xi$.
Alphabets will always be assumed to be finite and unordered.
Note that, if $\Sigma$~is unordered, so is $\bbT^\times_\xi\Sigma$
and $\bbU\bbT^\times_\xi\Sigma$ is just the power set.
Hence, we can regard every language $L \subseteq \bbT^\times_\xi\Sigma$
as an element of $\bbU\bbT^\times_\xi\Sigma$.

We aim for a characterisation of which elements of this set are regular languages.
Towards this goal we introduce a few operations on $\bbU\bbT^\times\Sigma$.
They are based on the well-known version for finite trees (see, e.g,. Section~2.4
of~\cite{LoedingThomas21}), suitably modified to work in the sorted setting
and to generate infinite trees.

Before presenting the definition we need to deal with the problem that
$\bbU \circ \bbT^\times$ does not form a monad and that
$\bbU\bbT^\times\Sigma$ not a $\bbU\bbT^\times$-algebra.
For this reason we will work with what we call \emph{bialgebras\?:}
a set~$A$ equipped both with a $\bbT^\times$-algebra product
$\pi : \bbT^\times A \to A$ and a $\bbU$-algebra product $\rho : \bbU A\to A$
(without any compatibility condition between them).
(Note that this is not the usual use of the word `bialgebra'.)
By Theorem~\ref{Thm: UTA is T-algebra}, $\bbU\bbT^\times\Sigma$ forms a bialgebra
with respect to the monads $\bbT^\times$~and~$\bbU$.

We use the following operations for our version of regular expressions\?:
\begin{itemize}
\item variables $x \in X$,
\item letters of the alphabet $a \in \Sigma$,
\item substitution~${}\cdot_x{}$, iteration~${-}^{+x}$, and $\omega$-power~${-}^{\omega x}$
  with respect to a single variable~$x$,
\item relabelling~${}^\sigma{-}$ of the variables,
\item union~$+$ and the empty language~$\emptyset$.
\end{itemize}
The formal definition is as follows.
\begin{defi}
Given a bialgebra $\frakA = \langle A,\pi,\rho\rangle$ we define the following operations.

(a) Each $a \in A_\xi$, induces an operation $a : A^\xi \to A$ by
\begin{align*}
  a(\bar b) := \pi(s)\,,
\end{align*}
where $s \in \bbT^\times_\xi A$ is the tree obtained form $\sing(a)$ by replacing
each leaf with label $x \in \xi$ by the tree~$\sing(b_x)$.

(b) For sorts $\xi,\zeta \in \Xi$ and a variable $x \in \xi$, we define a binary
\emph{substitution operation}
\begin{align*}
  {}\cdot_x{} : A_\xi \times A_\zeta \to A_{(\xi \setminus \{x\}) \cup \zeta}
  \qtextq{by}
  a \cdot_x b := \pi(s)\,,
\end{align*}
where $s$~is the tree obtained from $\sing(a)$ by replacing the leaf labelled~$x$
by the tree $\sing(b)$.

(c) For $a \in A_\xi$ and a surjective map $\sigma : \xi \to \zeta$, we set
\begin{align*}
  {}^\sigma a := \pi(s)\,,
\end{align*}
where $s$~is the tree obtained from~$\sing(a)$ by replacing each label $x \in \xi$
by $\sigma(x)$.

(d) We define ${+} : A_\xi \times A_\xi \to A_\xi$ and $\emptyset \in A_\xi$ by
\begin{align*}
  a + b := \rho(\Aboveseg\{a,b\})
  \qtextq{and}
  \emptyset := \rho(\emptyset)\,.
\end{align*}

(e)
Let $\zeta \in \Xi$. We call a tree~$s$ \emph{$\zeta$-trivial} if,
for all $v \in \dom(s)$ and $z \in \zeta$, we have
\begin{align*}
  s(v) = z
  \quad\iff\quad
  v \text{ is an $z$-successor.}
\end{align*}
(I.e., all $z$-successors are labelled by~$z$ and there are no other occurrences of~$z$.)
For a finite sequence of elements $a_i \in A_{\xi_i}$, $i < n$, and
a variable $x \in \zeta := \xi_0\cup\dots\cup \xi_{n-1}$, we
define the \emph{$\omega$-power} and the \emph{iteration} by
\begin{align*}
  &(a_0 +\dots+ a_{n-1})^{\omega x} := {} \\
  & \qquad\rho\bigl(\bigset{ \pi(s) }
      { s \in \bbT^\times_{\zeta \setminus \{x\}} \{a_0,\dots,a_{n-1}\}
        \text{ is $(\zeta \setminus \{x\})$-trivial } }\bigr)\,, \\[1ex]
  &(a_0 +\dots+ a_{n-1})^{+x} := {} \\
  & \qquad\begin{aligned}[t]
      \rho\bigl(\biglset \pi(s) \bigmset
          s \in \bbT^\times_\zeta \{a_0,\dots,a_{n-1}\} \text{ has finite height and it is}
              \quad & \\
          \text{$(\zeta \setminus \{x\})$-trivial } & \bigrset\bigr)\,.
    \end{aligned}
\end{align*}

(f) For a sort $\xi \in \Xi$ and a set~$\Sigma$,
the set $\bbE_\xi\Sigma$ of \emph{regular expression} over~$\Sigma$
consists of all finite terms~$R$ that can be built up from variables and
the operations (a)--(e) (for the bialgebra~$\bbU\bbT^\times\Sigma$), where
\begin{itemize}
\item we restrict the operations from~(a) to those where $a = \Aboveseg\sing(c)$,
  for some $c \in \Sigma$, and
\item the free variables are exactly those in~$\xi$.
\end{itemize}
We write $\lsem R\rsem \subseteq \bbU\bbT^\times\Sigma$ for the value of $R \in \bbE\Sigma$
in $\bbU\bbT^\times\Sigma$.
\markenddef
\end{defi}
\begin{rem}
The iteration and the $\omega$-power in~(e) have a built-in sum operation in order to support
choices between terms of different sorts, which is not possible using the normal sum
operation from~(d).
\markenddef
\end{rem}

\begin{exa}
We consider the alphabet $\Sigma = \{a,b,c\}$ where $a$~and~$b$ have sort $\{x,y\}$
and $c$~has sort~$\emptyset$.

(a) A regular expression for the language $\bbT^\times\Sigma$ is
\begin{align*}
  E := \bigl(\bigl(a(x,y) + b(x,y) + c\bigr)^{\omega x}\bigr)^{\omega y}.
\end{align*}

(b) An expression for the language of all trees with an infinite branch
labelled by~$a$ is given by
\begin{align*}
  R := \bigl(a(x,z) + a(z,y)\bigr)^{\omega z} \cdot_x E \cdot_y E\,.
\end{align*}

(c) Finally, the following expression describes all trees containing the letter~$a$.
\begin{align*}
  S := a(x,y)  \cdot_x E \cdot_y E
    + \bigl(b(x,z) + b(z,y)\bigr)^{+ z} \cdot_z a(x,y) \cdot_x E \cdot_y E\,.
\end{align*}
\upqed
\markenddef
\end{exa}

We still have to show that regular expressions capture the class of regular languages.
For the proof, let us quickly recall the notion of a tree automaton
(see, e.g., \cite{Thomas97,GraedelThomasWilke02,Loeding21} for details).
A~\emph{parity automaton} $\calA = \langle Q,\Sigma,\zeta,\Delta,q_\rmI,\Omega\rangle$
consists of a finite set~$Q$ of states, an input alphabet~$\Sigma$,
an input sort $\zeta \in \Xi$, an initial state $q_\rmI \in Q$, a priority function~$\Omega$,
and a transition relation
\begin{align*}
  \Delta \subseteq \sum_{\xi \in \Xi} (Q \times \Sigma_\xi \times Q^{\abs{\xi}})
                 + (Q \times \zeta)\,.
\end{align*}
A \emph{run}~$\rho$ of such an automaton on an input tree $t \in \bbT^\times_\zeta\Sigma$
is a labelling of~$t$ by states such that
\begin{itemize}
\item the root is labelled by~$q_\rmI$,
\item $\bigl\langle \rho(v),t(v),\rho(u_0),\dots,\rho(u_{n-1})\bigr\rangle \in \Delta$,
  for every vertex~$v$ with successors $u_0,\dots,u_{n-1}$,
\item every infinite branch $v_0,v_1,\dots$ of~$t$ satisfies the parity condition\?:
  \begin{align*}
    \liminf_{n \to \infty} \Omega(\rho(v_n)) \quad\text{is even.}
  \end{align*}
\end{itemize}
A \emph{partial run} is defined exactly like a run, except that the state at the root
can be arbitrary and that we do not require the transition relation to hold at vertices~$v$
labelled by a variable.
Let $\rho$~be a partial run on the tree $t \in \bbT^\times_\zeta\Sigma$.
The \emph{profile} of~$\rho$ is the pair $\langle p,(U_z)_{z \in \zeta}\rangle$
where $p$~is the state at the root and, for each variable $z \in \zeta$,
$U_z$~is the set of all pairs $\langle k,q\rangle$
such that there is a vertex~$v$ labelled~$z$ with state~$q$
and such that $k$~is the least priority seen along the path from the root to~$v$.
We define an ordering on profiles by
\begin{align*}
  \langle p,\bar U\rangle \leq \langle p',\bar U'\rangle
  \quad\defiff\quad
  p = p' \text{ and } U_z \subseteq U'_z \text{ for all } z \in \zeta\,.
\end{align*}
If $\sigma \leq \tau$, we say that the profile~$\sigma$ is \emph{bounded} by~$\tau$.

\begin{thm}
Let $\Sigma$~be an alphabet.
A~language $L \subseteq \bbT^\times_\zeta\Sigma$ is regular if, and only if,
$L = \lsem R\rsem$, for some regular expression $R \in \bbE_\zeta\Sigma$.
\end{thm}
\begin{proof}
$(\Leftarrow)$ The class of all regular tree languages
is closed under all operations that can appear in a regular expression.

$(\Rightarrow)$
Let $\calA = \langle Q,\Sigma,\zeta,\Delta,q_\rmI,\Omega\rangle$ be an automaton recognising~$L$
and fix an enumeration $q_0,\dots,q_{n-1}$ of~$Q$
such that $\Omega(q_0) \geq\dots\geq \Omega(q_{n-1})$.
For every profile~$\tau$ of~$\calA$ and every number $k \leq n$,
we will construct a regular expressions~$R^k_\tau$ defining the language
\begin{align*}
  \lsem R^k_\tau\rsem
    = \biglset t \in \bbT^\times\Sigma \bigmset {}
        &\text{there is a partial run on } t \text{ whose profile is bounded by} \\
        &\tau \text{ and whose internal states are among } q_0,\dots,q_{k-1} \bigrset\,.
\end{align*}
Then we obtain the desired expression for~$L$ by setting
\begin{align*}
  R := \sum_{\tau \in H} R^n_\tau\,,
\end{align*}
where $H$~is the set of all profiles $\tau = \langle q_\rmI,\bar U\rangle$ such that,
for all $z \in \zeta$,
\begin{align*}
  \langle k,p\rangle \in U_z
  \qtextq{implies}
  \langle p,z\rangle \in \Delta\,.
\end{align*}

We define the expressions $R^k_\tau$ by induction on~$k$.
For $k = 0$, we only need to consider runs without internal states. Hence, we can set
\begin{align*}
  R^0_\tau :=
    \sum {\lset a(\bar x) \mset {}}
        & a \in \Sigma\,, \text{ there is a partial run on } \sing(a) \text{ whose profile} \\
        & \text{is bounded by } \tau \rset\,.
\end{align*}
For the inductive step, suppose that $\tau = \langle p,\bar U\rangle$,
let $\xi$~be the sort of~$\tau$, and
let $D := \rng \Omega$ be the set of priorities used by~$\calA$.
We start with an expression describing runs starting with the state~$q_k$
and with only finitely many occurrences of~$q_k$ on each branch.
For a set $\eta \subseteq \xi$ of variables,
we write $\bar U|_\eta$ for the subtuple $(U_x)_{x \in \eta}$.
Let $V := D \times \{q_k\}$, let $y_0,y_1,\dots$ be new variables not in~$\xi$, and set
\begin{align*}
  T^k_{\bar U} := R^k_{q_k,\bar U} +
    \sum_{\zeta \cup \eta_0 \cup\dots\cup \eta_{n-1} = \xi}
      \begin{aligned}[t]
      &(S^{\zeta,n}_0 +\dots+ S^{\zeta,n}_{m-1})^{+y_0} \\
      &\quad{}\cdot_{y_0} R^k_{q_k,\bar U|_{\eta_0}}
              \cdot_{y_1}\cdots
              \cdot_{y_{n-1}} R^k_{q_k,\bar U|_{\eta_{n-1}}}\,,
     \end{aligned}
\end{align*}
where
\begin{itemize}
\item the sum ranges over all sequences $\zeta,\eta_0,\dots,\eta_{n-1}$ of subsets of~$\xi$
  whose union is equal to~$\xi$ and such that $\eta_i \neq \eta_j$, for $i \neq j$, and
\item $S^{\zeta,n}_0,\dots,S^{\zeta,n}_{m-1}$ is an enumeration of all expressions of the form
  $R^k_{q_k,\bar U|_\eta V\cdots V}$ where $\eta \subseteq \zeta$,
  $\upsilon \subseteq \{y_0,\dots,y_{n-1}\}$, and with $\abs{\upsilon}$~copies of~$V$
  that correspond to the variables $y \in \upsilon$.
\end{itemize}
Then $T^k_{\bar U}$ describes all trees that have a run with profile bounded by
$\langle q_k,\bar U\rangle$ and such that every branch contains only finitely many
occurrences of the state~$q_k$.

Similarly, we obtain an expression for all such trees with possibly infinitely many occurrences
of~$q_k$ by setting
\begin{align*}
  \hat T^k_{\bar U} := T^k_{\bar U} +
    \sum_{\zeta \cup \eta_0 \cup\dots\cup \eta_{n-1} = \xi}
      \begin{aligned}[t]
      &(S^{\zeta,n}_0 +\dots+ S^{\zeta,n}_{m-1})^{\omega z} \\
      &\quad{}\cdot_{y_0} T^k_{\bar U|_{\eta_0}}
              \cdot_{y_1}\cdots
              \cdot_{y_{n-1}} T^k_{\bar U|_{\eta_{n-1}}}\,,
      \end{aligned}
\end{align*}
where the $S^{\zeta,n}_i$~are defined as above, except that there is an additional copy of~$V$
corresponding to the variable~$z$.

If $\Omega(q_k)$~is odd, we can now set
\begin{align*}
  R^{k+1}_\tau :=
    R^k_\tau
    + \sum_{\zeta \cup \eta_0 \cup\dots\cup \eta_{n-1} = \xi}
         R^k_{p,\bar U|_\zeta V\dots V}
         \cdot_{y_0} T^k_{\bar U|_{\eta_0}}
         \cdot_{y_1}\cdots
         \cdot_{y_{n-1}} T^k_{\bar U|_{\eta_{n-1}}}\,.
\end{align*}
where the variables~$y_0,\dots,y_{n-1}$ are the ones corresponding to the $n$~copies of the
set~$V$.
If $\Omega(q_k)$~is even, we instead use
\begin{align*}
  R^{k+1}_\tau :=
    R^k_\tau
    + \sum_{\zeta \cup \eta_0 \cup\dots\cup \eta_{n-1} = \xi}
         R^k_{p,\bar U|_\zeta V\dots V}
         \cdot_{y_0} \hat T^k_{\bar U|_{\eta_0}}
         \cdot_{y_1}\cdots
         \cdot_{y_{n-1}} \hat T^k_{\bar U|_{\eta_{n-1}}}\,.
\end{align*}
\upqed
\end{proof}

\section{Conclusion}   

We have introduced the upwards-closed power-set monad~$\bbU$ on $\Pos^\Xi$
and studied possible distributive laws between it and two monads of infinite trees\?:
linear trees~$\bbT$ and non-linear ones~$\bbT^\times$.
For the monad~$\bbT$, we have shown in Theorems
\ref{Thm: distributive law for polynomial functors}~and~\ref{Thm: dist unique}
that there exists a unique distributive law $\dist : \bbT\bbU \Rightarrow \bbU\bbT$.
For the monad~$\bbT^\times$ on the other hand, we have proved in
Theorem~\ref{Thm: dist iff linear} that there is no distributive law
$\bbT^\times\bbU \Rightarrow \bbU\bbT^\times$.
Our main result (Theorem~\ref{Thm: UTA is T-algebra}) states that, nevertheless,
every set of the form~$\bbU\bbT^\times A$ forms a $\bbT^\times$-algebra when
equipped with a suitable product.
The two examples in Section \ref{Sect:substitutions}~and~\ref{Sect:regex} show
that this partial result is frequently sufficient for applications.

There are several possible directions where one can go from here.
Of interest to language theorists would be to consider other functors
similar to the power-set one, for instance the functor producing linear combinations
over a given semiring, or similar analogues of the power-set functor for weighted languages.

More category-theoretically inspired considerations would include a more systematic study
of when a distributive law with the power-set monad exists.
In particular, it would be interesting to transfer the results in Section~\ref{Sect:polynomial}
from polynomial monads to quotients of such monads.
Another avenue to pursue would be to generalise Theorem~\ref{Thm: dist iff linear}
to other monads than the power-set one
by extracting the abstract properties of the power-set monad needed for the proof.

{\small\raggedright
\bibliographystyle{alphaurl}
\bibliography{Power}}

\end{document}